\documentclass[aps,showkeys,preprintnumbers,superscriptaddress,floatfix,footinbib]{revtex4}

\usepackage{epsfig}
\usepackage{bm}
\usepackage{amssymb}
\usepackage{amsmath}
\usepackage{color}
\usepackage{subfigure}
\usepackage{dcolumn}

\usepackage[utf8]{inputenc}

\usepackage{amsmath}
\usepackage{amssymb}
\usepackage{amsthm}
\usepackage{amscd, amsfonts, mathrsfs}
\usepackage{cases}
\usepackage{verbatim} 
\usepackage{epsf}
\usepackage[bookmarksnumbered,pdfpagelabels=true,plainpages=false,colorlinks=true,
            linkcolor=black,citecolor=black,urlcolor=black]{hyperref}

\theoremstyle{plain}
\newtheorem{theorem}{Theorem}

\theoremstyle{definition}

\newtheorem{definition}[theorem]{Definition}

\allowdisplaybreaks

\newcommand{\Si}{\Sigma}

\newcommand{\be}{\begin{equation}}
\newcommand{\ee}{\end{equation}}
\newcommand{\bea}{\begin{eqnarray}}
\newcommand{\eea}{\end{eqnarray}}
\newcommand{\bml}{\begin{subequations}}
\newcommand{\eml}{\end{subequations}}

\newcommand{\bbm}{\begin{bmatrix}}
\newcommand{\ebm}{\end{bmatrix}}

\begin{document}


\title{Nonlinear Constraints on Relativistic Fluids Far From Equilibrium}
\date{\today}

\author{F\'abio S.\ Bemfica}
\affiliation{Escola de Ci\^encias e Tecnologia, Universidade Federal do Rio Grande do Norte, 59072-970, Natal, RN, Brazil}
\email{fabio.bemfica@ect.ufrn.br}

\author{Marcelo M.\ Disconzi}
\affiliation{Department of Mathematics, Vanderbilt University, Nashville, TN, USA}
\email{marcelo.disconzi@vanderbilt.edu}

\author{Vu Hoang}
\affiliation{Department of Mathematics, The University of Texas at San Antonio, One UTSA Circle, San Antonio, TX 78249, USA}
\email{duynguyenvu.hoang@utsa.edu}
\email{maria\_radosz@hotmail.com}

\author{Jorge Noronha}
\affiliation{Department of Physics, University of Illinois, 1110 W. Green St., Urbana IL 61801-3080, USA}
\email{jn0508@illinois.edu}

\author{Maria Radosz}
\affiliation{Department of Mathematics, The University of Texas at San Antonio, One UTSA Circle, San Antonio, TX 78249, USA}
\email{maria_radosz@hotmail.com}


\begin{abstract}
New constraints are found that must necessarily hold for Israel-Stewart-like theories of fluid dynamics to be causal far away from equilibrium. Conditions that are sufficient to ensure causality, local existence, and uniqueness of solutions in these theories are also presented. Our results hold in the full nonlinear regime, taking into account bulk and shear viscosities (at zero chemical potential), without any simplifying symmetry or near-equilibrium assumptions. Our findings provide fundamental constraints on the magnitude of viscous corrections 
in fluid dynamics far from equilibrium.
\end{abstract}



\maketitle


\textbf{1. Introduction.} Relativistic fluid dynamics is essential to the state-of-the-art modeling of the quark-gluon plasma (QGP) formed in ultrarelativistic heavy-ion collisions (see \cite{Heinz:2013th,Gale:2013da,Romatschke:2017ejr}). However, despite its wide use and significant success, it remains unclear why such a fluid dynamical description is applicable given that local deviations from equilibrium in nucleus-nucleus collisions can be very large, especially at early times \cite{Schenke:2012wb,Niemi:2014wta,Noronha-Hostler:2015coa}. In fact, typical fluid-like signatures involving anisotropic flow \cite{Luzum:2013yya} persist even in small systems formed in proton-nucleus and proton-proton collisions at sufficiently high multiplicity \cite{Bozek:2011if,Khachatryan:2015waa,Aad:2015gqa,Khachatryan:2016txc,Weller:2017tsr,PHENIX:2018lia,Acharya:2019vdf}. Such findings have motivated a series of new investigations on the foundations of relativistic viscous fluid dynamics \cite{BemficaDisconziNoronha,BemficaDisconziNoronhaNonconformalBarotropic,Kovtun:2019hdm,Hoult:2020eho} and their subsequent extension towards the far-from-equilibrium regime relevant for heavy-ion collisions \cite{Heller:2013fn,Heller:2015dha,Buchel:2016cbj,Denicol:2016bjh,Heller:2016rtz,Romatschke:2017vte,Spalinski:2017mel,Strickland:2017kux,Romatschke:2017acs,Florkowski:2017olj,Denicol:2017lxn,Behtash:2017wqg,Blaizot:2017ucy,Almaalol:2018ynx,Denicol:2018pak,Gallmeister:2018mcn,Casalderrey-Solana:2018uag,Behtash:2018moe,Behtash:2019txb,Strickland:2018ayk,Jaiswal:2019cju,Kurkela:2020sal,Giacalone:2019ldn,Denicol:2019lio,Chattopadhyay:2019jqj,Almaalol:2020rnu,Das:2020fnr}.  

The viscous fluid description of the QGP is currently based on ideas from Israel and Stewart (IS) \cite{MIS-2,MIS-6} (see also Mueller \cite{MIS-1}), who proposed a way to fix the long-standing acausality \cite{PichonViscous}  and instability \cite{Hiscock_Lindblom_instability_1985} problems of the relativistic generalization of Navier-Stokes (NS) equations derived by Eckart \cite{EckartViscous} and Landau and Lifshitz \cite{LandauLifshitzFluids}. The general mechanism introduced by IS to try to avoid such issues assumes that dissipative currents such as the shear stress tensor, $\pi_{\mu\nu}$, and the bulk scalar, $\Pi$, obey nonlinear relaxation equations describing how such quantities relax to their relativistic NS limits within relaxation time scales $\tau_\pi$ and $\tau_\Pi$. The same principle is also at play in modern formulations of fluid dynamics put forward by Ref.\
 \cite{Baier:2007ix} and Ref.\ \cite{Denicol:2012cn}, which are currently employed in numerical simulations (see, for instance, \cite{Ryu:2017qzn}). 

It is well-known that the IS-like theories are linearly stable around equilibrium \cite{Hiscock_Lindblom_stability_1983,Olson:1989ey,Denicol:2008ha,Pu:2009fj}. But physically sensible \emph{relativistic} theories of fluid dynamics must also be causal, i.e., the equations of motion must be hyperbolic and the propagation of information must be at most the speed of light \cite{WaldBookGR1984}. Also, the Cauchy problem must be locally well-posed \cite{ChoquetBruhatGRBook}, i.e., given initial conditions one must show that
the equations admit a unique solution. A common \emph{misconception} in the field is that IS-like theories have already been proven to be causal a long time ago in Refs.\ \cite{Hiscock_Lindblom_stability_1983,Olson:1989ey}. This is \emph{not} the case. Those early works \emph{only} considered
linearized disturbances around equilibrium, where the background fields $\pi_{\mu\nu}$ and $\Pi$ 
vanish and the corresponding linear disturbances are small.
Such a linearized analysis says nothing about the nonlinear regime, \emph{even for small}
$\pi_{\mu\nu}$ and $\Pi$. The far-from-equilibrium regime, in particular, is necessarily 
nonlinear as $\pi_{\mu\nu}$ and $\Pi$ can be as large as the local equilibrium pressure $P$. 

Hence, it is not known if IS-theories are indeed sensible in the regime probed by high energy hadronic collisions. This key question must be answered to ensure that general conclusions regarding the formation of the QGP (e.g., in proton-proton collisions) are sensible. 
Understanding the far-from-equilibrium properties of such theories is also crucial to reliably assess the role of
viscous effects in early universe cosmology \cite{Brevik:2017msy}.
Here, we make essential steps towards solving this critical problem by finding conditions that
\emph{must necessarily hold} for IS-like theories to be causal. We also present conditions that are \emph{sufficient to ensure} causality, local existence, and uniqueness of solutions of IS-like theories.
Our results hold in the full nonlinear regime, with bulk and shear viscosities (at zero chemical potential),
in three spatial dimensions, without any symmetry or near-equilibrium assumptions.
Our conditions are simple algebraic inequalities that
can be easily checked in a given problem.
This is the first time that such general statements (causality, local existence, uniqueness) are proven for IS-like theories with shear and bulk viscosities in the full nonlinear regime without 
simplifying dynamical assumptions.


\vskip 0.1cm
\textbf{2. The equations of motion.} Using the Landau frame definition of the hydrodynamic variables \cite{LandauLifshitzFluids}, the energy-momentum tensor of the fluid can be written as \footnote{We use units $c = \hbar = k_B = 1$. The space-time metric signature is $(-+++)$. Greek indices run from 0 to 3, Latin indices from 1 to 3.}
$
T^{\mu\nu} = \varepsilon u^\mu u^\nu + (P+\Pi)\Delta^{\mu\nu} + \pi^{\mu\nu},
$
where $u^\mu$ is the fluid's 4-velocity (with $u_\mu u^\mu = -1$), $\varepsilon$ is the energy density,
$P = P(\varepsilon)$ is the equilibrium pressure defined by an equation of state, $\Delta_{\mu\nu} = g_{\mu\nu}+u_\mu u_\nu$ is the projector orthogonal to the flow, $g_{\mu\nu}$ is the spacetime metric, $\pi_{\mu\nu}=\pi_{\nu\mu}$, $\pi^{\mu\nu}u_\mu = 0$, and $\Delta_{\mu\nu}\pi^{\mu\nu} = 0$. We focus on high energy collisions and, thus, we only investigate here the case of zero chemical potentials. Conservation of energy and momentum implies that $\nabla_\mu T^{\mu\nu} = 0$, 
which can be written as ($c_s^2 = dP/d\varepsilon$ is the equilibrium speed of sound squared)
\bea
&&u^\alpha \nabla_\alpha \varepsilon + \left(\varepsilon+P+\Pi\right)\nabla_\alpha u^\alpha +\pi^{\alpha}_{\mu}\nabla_\alpha u^\mu= 0,\nonumber\\
&&\left(\varepsilon+P +\Pi\right)u^\beta \nabla_\beta u_\alpha + c_s^2\Delta^{\beta}_\alpha \nabla_\beta \varepsilon + \Delta^{\beta}_\alpha \nabla_\beta\Pi+ \Delta_\alpha^\beta \nabla_\mu \pi^{\mu}_{\beta}=0.\label{conservaEM}
\eea
Here, we consider the case where the dissipative currents $\{\pi^{\mu\nu},\Pi\}$ 
satisfy the following equations \footnote{Note that our metric signature is different than in \cite{Denicol:2012cn}.}, derived using the DNMR formalism \cite{Denicol:2012cn}, and commonly used in heavy-ion collision applications,
\bml
\label{supplemental}
\bea
\tau_\Pi u^\mu \nabla_\mu \Pi + \Pi &=& -\zeta \nabla_\mu u^\mu -\delta_{\Pi\Pi}\Pi \nabla_\mu u^\mu -\lambda_{\Pi \pi} \pi^{\mu\nu}\sigma_{\mu\nu},\label{bulk}\\
 \tau_\pi \Delta^{\mu\nu}_{\alpha\beta}u^\lambda \nabla_\lambda \pi^{\alpha\beta} + \pi^{\mu\nu} &=& -2 \eta \sigma^{\mu\nu} -\delta_{\pi \pi} \pi^{\mu\nu}\nabla_\alpha u^\alpha  -\tau_{\pi\pi}\pi_\alpha^{\langle \mu}\sigma^{\nu\rangle\alpha} - \lambda_{\pi\Pi}\Pi \sigma^{\mu\nu},
\eea
\eml
where $\sigma^{\mu\nu} = \Delta^{\mu\nu}_{\alpha\beta}\nabla^\alpha u^\beta$ is the shear tensor, 
$\Delta^{\mu\nu}_{\alpha\beta} = \left(\Delta^{\mu}_\alpha \Delta^\nu_\beta + \Delta^{\mu}_\beta \Delta^\nu_\alpha\right)/2 - \frac{1}{3}\Delta^{\mu\nu} \Delta_{\alpha\beta}$, $A_\lambda^{\langle \mu}B^{\nu\rangle\lambda} =  \Delta^{\mu\nu}_{\alpha\beta}A^{\alpha \lambda}B^\beta_\lambda$, and $\eta$, $\zeta$ are the shear and bulk viscosities, respectively. All the transport coefficients, $\{\eta,\zeta, \tau_\Pi, \tau_\pi,\delta_{\Pi\Pi},\lambda_{\Pi \pi},\delta_{\pi \pi},\tau_{\pi\pi},\lambda_{\pi\Pi}\}$, can depend on the ten dynamical variables $\{\varepsilon,u_\mu,\pi_{\mu\nu},\Pi\}$ (so, in principle, they may even depend on the dissipative tensors) but not on their derivatives.
Explicit expressions for transport coefficients in models can be found, for instance, in \cite{Denicol:2012cn,Denicol:2014vaa,Finazzo:2014cna}. 

We note that $\{\eta,\zeta,\tau_\pi,\tau_\Pi\}$ are the only coefficients that remain after linearization around equilibrium where $\pi^{\mu\nu}=0$ and $\Pi=0$. This shows why linearized analyses \cite{Hiscock_Lindblom_stability_1983,Olson:1989ey} necessarily miss the effects from the other coefficients, $\{\delta_{\Pi\Pi},\lambda_{\Pi \pi},\delta_{\pi \pi},\tau_{\pi\pi},\lambda_{\pi\Pi}\}$, which contribute to the nonlinear evolution. However, other nonlinear terms such as $\pi_{\mu\nu}\pi^{\mu\nu}$, $\Pi^2$, $\pi^{\mu\nu}\Pi$, $\pi_\alpha^{\langle \mu}\pi^{\nu\rangle\alpha}$, which appear in \cite{Denicol:2012cn}, could have been trivially added to the equations as they do not contribute
to a causality analysis since they do not involve derivatives of the fields. Nevertheless, there are still some other nonlinear terms that can be considered such as $\pi_\alpha^{\langle \mu}\Omega^{\nu\rangle\alpha}$, where $\Omega_{\mu\nu} = (\Delta_{\mu}^\alpha \nabla_\alpha u_\nu - \Delta_{\nu}^\alpha \nabla_\alpha u_\mu)/2$ is the vorticity, and also $\Omega_\alpha^{\langle \mu}\Omega^{\nu\rangle\alpha}$ \cite{Romatschke:2017ejr}. The former will be investigated in a separate publication. The latter contributes with derivatives of the fields to the principal part of the system of equations and, thus, a different analysis than presented here would be required.


\vskip 0.1cm
\textbf{3. Causality.} Causality is the concept in relativity theory asserting that no information propagates faster than the speed of light and no closed timelike curves exist (so the future cannot influence 
the past). See the Supplemental Material and references \cite{DisconziViscousFluidsNonlinearity,
KatoQuasiLinear,Fischer_Marsden_Einstein_FOSH_1972,EvansPDE,
MajdaCompressibleFlow}
for a mathematically precise definition of causality. 
Causality can be investigated by determining the characteristic manifolds associated with a system of PDE's. In fact, the existence of domains of dependence 
for solutions of a system of PDEs, as well as their corresponding propagation speeds,
can be inferred from the system's characteristics \cite{Leray_book_hyperbolic,DisconziSpeckRelEulerNull}.
Let us write equations \eqref{conservaEM}-\eqref{supplemental} as  $A^\alpha\nabla_\alpha\Psi=F(\Psi)$, where we defined the vector $\Psi=(\varepsilon,u^\nu,\Pi,\pi^{0\nu},\pi^{1\nu},\pi^{2\nu},\pi^{3\nu})$,
the $22\times22$ matrix 
\be
\label{E:Matrix}
A^\alpha=\bbm
u^\alpha & \rho\delta^\alpha_\nu+\pi^\alpha_\nu & 0_{1\times 1} & 0_{1\times 4} & 0_{1\times 4} & 0_{1\times 4} & 0_{1\times 4}\\
c_s^2\Delta^{\mu\alpha} & \rho u^\alpha\delta^\mu_\nu -\pi^\alpha_\nu u^\mu & \Delta^{\mu\alpha} & \delta^\alpha_0 I_4 & \delta^\alpha_1 I_4 & \delta^\alpha_2 I_4 & \delta^\alpha_3 I_4\\
0_{4\times 1} & E^\alpha_\nu & \tau_\Pi u^\alpha & 0_{4\times 4} & 0_{4\times 4} & 0_{4\times 4} & 0_{4\times 4}\\
0_{4\times 1} & C^{0\delta\alpha}_\nu & 0_{4\times 1} & \tau_\pi u^\alpha I_4 & 0_{4\times 4} & 0_{4\times 4} & 0_{4\times 4}\\
0_{4\times 1} & C^{1\delta\alpha}_\nu & 0_{4\times 1} & 0_{4\times 4} & \tau_\pi u^\alpha I_4 & 0_{4\times 4} & 0_{4\times 4}\\
0_{4\times 1} & C^{2\delta\alpha}_\nu & 0_{4\times 1} & 0_{4\times 4} & 0_{4\times 4} & \tau_\pi u^\alpha I_4 & 0_{4\times 4}\\
0_{4\times 1} & C^{3\delta\alpha}_\nu & 0_{4\times 1} & 0_{4\times 4} & 0_{4\times 4} & 0_{4\times 4} & \tau_\pi u^\alpha I_4
\ebm,
\ee
and $F(\Psi)$ is a vector that does not contain derivatives of the variables. Above, we also defined 
$\rho =\varepsilon+P+\Pi$, $E^\alpha_\nu=\left (\zeta+\delta_{\Pi\Pi}\Pi\right )\delta^\alpha_\nu+\lambda_{\Pi \pi} \pi^{\alpha}_\nu$, $B^{\mu\lambda\alpha}_\nu=\dfrac{1}{2}\left (\Delta^{\mu\alpha}\delta^\lambda_\nu+\Delta^{\lambda\alpha}\delta^\mu_\nu-\frac{2}{3}\Delta^{\mu\lambda}\delta^\alpha_\nu\right )$, and 
\bea
C^{\sigma\delta\alpha}_\nu&=&\left [(2 \eta+\lambda_{\pi\Pi}\Pi)\delta^\sigma_\mu\delta^\delta_\lambda+\frac{\tau_{\pi\pi}}{2}\pi_\lambda^{\sigma}\delta^{\delta}_\mu+\frac{\tau_{\pi\pi}}{2}\pi_\lambda^{\delta}\delta^{\sigma}_\mu\right ]B^{\mu\lambda\alpha}_\nu-\frac{\tau_{\pi\pi}}{3}\Delta^{\sigma\delta}\pi^{\alpha}_\nu+\delta_{\pi \pi} \pi^{\sigma\delta}\delta^\alpha_\nu\nonumber\\
&&-\tau_\pi(\pi^\sigma_\nu u^\delta+\pi^\delta_\nu u^\sigma)u^\alpha.
\nonumber
\eea
The characteristic surfaces $\{ \Phi(x) = 0 \}$ are determined by the principal part of the equations by solving the characteristic equation 
$\det(A^\alpha \xi_\alpha)=0$, with $\xi_\alpha = \nabla_\alpha \Phi$ \cite{Courant_and_Hilbert_book_2}.
The system is causal 
if, for any $\xi_i$, it holds that (C1) the roots $\xi_0 = \xi_0(\xi_i)$ of the characteristic equation
are real (in particular, the system will be hyperbolic) and (C2) $\xi_\alpha = (\xi_0(\xi_i),\xi_i)$
is spacelike or 
lightlike.
Condition (C2) implies that the characteristic surfaces $\{ \Phi(x) = 0\}$ are timelike or lightlike, 
indicating that no information is superluminal. 

From \eqref{E:Matrix}, it is clear that the characteristics associated with
the evolution depend on the dissipative tensors $\{\pi^{\mu\nu},\Pi\}$. Therefore, the true causal behavior 
of IS-theories is necessarily a far-from-equilibrium property of the fluid 
and linear analyses around equilibrium cannot be used to establish causality and well-posedness in IS-theories. The computation of the characteristics defined by \eqref{E:Matrix}, which is needed
for a causality analysis, is extremely involved and 
is presented in the Supplemental Material.

Let $\Lambda_\alpha$, $\alpha=0,1,2,3$, be the eigenvalues of the $\pi^\mu_\nu$.
The eigenvalues are such that $\Lambda_0=0$, since $u_\mu$ is in the kernel of
$\pi^\mu_\nu$ ($u_\mu \pi^\mu_\nu=0$), and $\Lambda_1+\Lambda_2+\Lambda_3=0$,
so that the trace is kept zero.
Without loss of generality, let us take $\Lambda_1\le \Lambda_2\le \Lambda_3$ with $\Lambda_1\le 0 \le \Lambda_3$. We now state our assumptions, which are the following:
(A1) for the transport coefficients and relaxation times, 
suppose that $\tau_\Pi, \tau_\pi>0$ and $\eta,\zeta, \tau_{\pi\pi},\delta_{\Pi\Pi},\lambda_{\Pi \pi},\delta_{\pi \pi},\lambda_{\pi\Pi},c_s^2\ge0$; (A2) for the fluid variables, suppose that
$\varepsilon >0$, $P \geq 0$, and $\varepsilon+P +\Pi> 0$;
finally, also assume that  (A3) $\varepsilon+P +\Pi + \Lambda_a > 0$, $a=1,2,3$.
Then, the following conditions are \emph{necessary} for causality,
i.e., \emph{if any of the inequalities below is not satisfied then the system is not causal:}
\bml
\label{necessary_conditions}
\bea
&&(2 \eta+\lambda_{\pi\Pi}\Pi)-\frac{1}{2}\tau_{\pi\pi}|\Lambda_1| \geq 0\\
&&\varepsilon+P+\Pi-\frac{1}{2\tau_\pi}(2 \eta+\lambda_{\pi\Pi}\Pi)-\frac{\tau_{\pi\pi}}{4\tau_\pi}\Lambda_3\ge 0,\label{n12}\\
&&
\frac{1}{2\tau_\pi}(2 \eta+\lambda_{\pi\Pi}\Pi)+\frac{\tau_{\pi\pi}}{4\tau_\pi}\left (\Lambda_a +\Lambda_d\right ) \geq 0,\quad a\ne d,
\label{new_necessary_condition}
\\
&&\varepsilon+P+\Pi+\Lambda_a-\frac{1}{2\tau_\pi}(2 \eta+\lambda_{\pi\Pi}\Pi)-\frac{\tau_{\pi\pi}}{4\tau_\pi}\left (\Lambda_d +\Lambda_a\right )\ge0,\quad a\ne d\label{n11}\\
&& \frac{1}{2\tau_\pi}(2 \eta+\lambda_{\pi\Pi}\Pi)+\frac{\tau_{\pi\pi}}{2\tau_\pi}\Lambda_d
+\frac{1}{6\tau_\pi}[2 \eta+\lambda_{\pi\Pi}\Pi+(6\delta_{\pi\pi}-\tau_{\pi\pi})\Lambda_d]
\nonumber \\
&& +\frac{\zeta+\delta_{\Pi\Pi}\Pi+\lambda_{\Pi \pi}\Lambda_d}{\tau_\Pi}+(\varepsilon+P+\Pi+\Lambda_d)c_s^2 \geq 0,
\label{n1}\\
&& \varepsilon+P+\Pi+\Lambda_d -\frac{1}{2\tau_\pi}(2 \eta+\lambda_{\pi\Pi}\Pi)-\frac{\tau_{\pi\pi}}{2\tau_\pi}\Lambda_d
-\frac{1}{6\tau_\pi}[2 \eta+\lambda_{\pi\Pi}\Pi+(6\delta_{\pi\pi}-\tau_{\pi\pi})\Lambda_d]
\nonumber\\
&&-\frac{\zeta+\delta_{\Pi\Pi}\Pi+\lambda_{\Pi \pi}\Lambda_d}{\tau_\Pi}-(\varepsilon+P+\Pi+\Lambda_d)c_s^2 \geq 0,
\label{n2}\eea
\eml
where \eqref{new_necessary_condition}-\eqref{n2} must hold for $a,d=1, 2,3$. 
The proof that \eqref{necessary_conditions} are necessary conditions for causality
under assumptions (A1)-(A3) is given in the Supplemental Material. Here, we discuss the significance 
of this result.

We stress that assumptions (A1) and (A2) are standard in heavy-ion collision applications \cite{Ryu:2017qzn},
and (A3) is a very natural assumption since $P +\Pi + \Lambda_a$ for $a=1,2,3$ may be interpreted as the pressure in each spatial axis in the local rest frame. Furthermore, it is natural to make assumptions that
hold close to equilibrium, and since (A2) guarantees $\varepsilon+P +\Pi > 0$, for small deviations from equilibrium $\Lambda_a$ will be small, giving
$\varepsilon+P +\Pi + \Lambda_a > 0$. That said, we stress that although (A3) is expected to hold
near equilibrium, it is itself \emph{not} a near-equilibrium assumption.

Conditions \eqref{necessary_conditions}
could never have been found using a linearized analysis as they depend on $\Pi$ and $\Lambda_a$ both of which vanish in equilibrium. Consequently, if in any fluid dynamic simulation in heavy-ion collisions that employs \eqref{conservaEM}-\eqref{supplemental} the necessary conditions above are not fulfilled, causality is necessarily violated. It is important to point out that this causality
violation has nothing to do with the ability of numerical schemes to produce a solution,
a point we discuss in the Conclusion. 

While the above conditions must hold for the system to be causal, 
they are not sufficient conditions, i.e., by themselves, conditions (A1)-(A3) and \eqref{necessary_conditions} 
do not assure the system to be causal (see the Supplemental Material).  
Therefore it is important to have
conditions that are sufficient for causality.
In this regard, assume again that (A1)-(A3) hold.
Then the following conditions are \emph{sufficient} to ensure that causality holds, i.e., \emph{if they are satisfied then the system is causal:}
\bml
\label{conditions}
\bea
&&(\varepsilon+P+\Pi-|\Lambda_1|)-\frac{1}{2\tau_\pi}(2 \eta+\lambda_{\pi\Pi}\Pi)-\frac{\tau_{\pi\pi}}{2\tau_\pi}\Lambda_3\ge 0,\label{cond1-1}\\
&&(2 \eta+\lambda_{\pi\Pi}\Pi)-\tau_{\pi\pi}|\Lambda_1|>0,\label{cond1-2}\\
&&\tau_{\pi\pi}\le 6\delta_{\pi\pi},\label{cond4a}\\
&&\frac{\lambda_{\Pi \pi} }{\tau_\Pi}+c_s^2-\frac{\tau_{\pi\pi}}{12\tau_\pi}\ge 0,\label{cond4b}\\
&&\frac{1}{3\tau_\pi}[4 \eta+2\lambda_{\pi\Pi}\Pi+(3\delta_{\pi\pi}+\tau_{\pi\pi})\Lambda_3]+\frac{\zeta+\delta_{\Pi\Pi}\Pi+\lambda_{\Pi \pi}\Lambda_3}{\tau_\Pi}+|\Lambda_1|+\Lambda_3 c_s^2 \nonumber\\
&&+\frac{\frac{12\delta_{\pi\pi}-\tau_{\pi\pi}}{12\tau_\pi}\left (\frac{\lambda_{\Pi \pi} }{\tau_\Pi}+c_s^2-\frac{\tau_{\pi\pi}}{12\tau_\pi}\right )(\Lambda_3+|\Lambda_1|)^2}{\varepsilon+P+\Pi-|\Lambda_1|-\frac{1}{2\tau_\pi}(2 \eta+\lambda_{\pi\Pi}\Pi)-\frac{\tau_{\pi\pi}}{2\tau_\pi}\Lambda_3}\le (\varepsilon+P+\Pi)(1-c_s^2),\label{cond5}\\
&&\frac{1}{6\tau_\pi}[2 \eta+\lambda_{\pi\Pi}\Pi+(\tau_{\pi\pi}-6\delta_{\pi\pi})|\Lambda_1|]+\frac{\zeta+\delta_{\Pi\Pi}\Pi-\lambda_{\Pi \pi}|\Lambda_1|}{\tau_\Pi}+(\varepsilon+P+\Pi-|\Lambda_1|)c_s^2 \ge 0,\label{cond7}\\
&&1\ge\frac{\frac{12\delta_{\pi\pi}-\tau_{\pi\pi}}{12\tau_\pi}\left (\frac{\lambda_{\Pi \pi} }{\tau_\Pi}+c_s^2-\frac{\tau_{\pi\pi}}{12\tau_\pi}\right )(\Lambda_3+|\Lambda_1|)^2}{\left [\frac{1}{2\tau_\pi}(2 \eta+\lambda_{\pi\Pi}\Pi)-\frac{\tau_{\pi\pi}}{2\tau_\pi}|\Lambda_1|\right ]^2}\label{cond6}\\
&&\frac{1}{3\tau_\pi}[4 \eta+2\lambda_{\pi\Pi}\Pi-(3\delta_{\pi\pi}+\tau_{\pi\pi})|\Lambda_1|]+\frac{\zeta+\delta_{\Pi\Pi}\Pi-\lambda_{\Pi \pi}|\Lambda_1|}{\tau_\Pi}+(\varepsilon+P+\Pi-|\Lambda_1|)c_s^2\nonumber\\
&&\ge \frac{(\varepsilon+P+\Pi+\Lambda_2)(\varepsilon+P+\Pi+\Lambda_3)}{3(\varepsilon+P+\Pi-|\Lambda_1|)}\left \{1+\frac{2\left [\frac{1}{2\tau_\pi}(2 \eta+\lambda_{\pi\Pi}\Pi)+\frac{\tau_{\pi\pi}}{2\tau_\pi}\Lambda_3\right ]}{\varepsilon+P+\Pi-|\Lambda_1|}\right \},\label{cond8}
\eea
\eml
where condition \eqref{cond8} can be dropped if $\delta_{\pi\pi}=\tau_{\pi\pi}=0$. 
The detailed proof can be found in the Supplemental Material. 
Since \eqref{necessary_conditions} must hold for causality, they must be satisfied
for any set of conditions that imply causality, and it is possible to verify that 
\eqref{conditions} imply \eqref{necessary_conditions} under assumptions
(A1)-(A3). When shear viscous effects are neglected, \eqref{conditions} reduces to the conditions for the bulk viscosity case found in \cite{BemficaDisconziNoronha_IS_bulk}.

Conditions (A1)-(A3)-\eqref{conditions} also ensure the unique local solvability of the initial-value problem
in the class of quasi-analytic functions. More precisely, given 
initial data of sufficient regularity satisfying \eqref{conditions}, there exists a unique solution
to the nonlinear equations taking the given initial data, defined for a certain time interval 
(again, we refer to the Supplemental Material for details). Therefore, if (A1)-(A3) and \eqref{conditions} hold, the evolution of the viscous fluid is guaranteed to be well defined and  causal even far from equilibrium where the gradients (and, hence, $\pi^{\mu\nu}$ and $\Pi$) are large. This is especially relevant for the open question in heavy-ion collisions concerning the properties of hydrodynamic attractors \cite{Heller:2015dha} under general flow conditions \cite{Romatschke:2017acs,Denicol:2020eij} and also for an overall validation of a fluid dynamic description of small systems, such as proton-proton collisions.  

Although here we focus on applications to heavy-ion collisions,
where $g_{\mu\nu}$ is the Minkowski metric, it is not difficult to see
that the methods of \cite{BemficaDisconziNoronha_IS_bulk} can be adapted to show
that our conclusions hold when \eqref{conservaEM}-\eqref{supplemental} are coupled to Einstein's
equations (see the Supplemental Material).  Therefore, our results are also crucial to determine the far from equilibrium behavior of viscous fluids with shear and bulk viscosity in general relativity, which may be directly relevant to neutron star mergers \cite{RezzollaImportanceViscosityNeutronStars}.  

When we linearize the equations around the equilibrium, terms involving 
 $\tau_{\pi\pi},\delta_{\Pi\Pi},\lambda_{\Pi \pi},\delta_{\pi \pi},\lambda_{\pi\Pi}$ drop out and, thus, (A1) can be replaced by $ \tau_\pi, \tau_\Pi >0$, $\eta,\zeta, c_s^2 \geq 0$ and
 (A2) and (A3) can be replaced by $\varepsilon + P >0$ and $P\geq 0$.
Then, conditions \eqref{conditions} become necessary and reduce to  $\varepsilon+P>0$, $\varepsilon+P-\frac{\eta}{\tau_\pi}\ge 0$ and $\frac{1}{\varepsilon+P}\left (\frac{4\eta}{3\tau_\pi}+\frac{\zeta}{\tau_\Pi}\right ) \le 1-c_s^2$. These conditions coincide with the corresponding well-known results previously found in \cite{Hiscock_Lindblom_stability_1983,Olson:1989ey} that ensure causality and stability in the linearized regime around equilibrium.

We presented two sets of conditions for causality, namely, conditions that are
necessary and conditions that are sufficient. Further studies must be done to discover conditions that
are necessary \emph{and} sufficient, i.e., conditions that ensure the system to be causal \emph{if and only if} they
hold. This is an extremely challenging task given the complexity of the characteristic equation in the nonlinear problem, and would require developing  essential new ideas to analyze its roots. 
 

\vskip 0.1cm
\textbf{4. Conformal limit.} As an application of our nonlinear constraints, consider 
a conformal fluid \cite{Baier:2007ix}, i.e., $\Pi=0$, $P = \varepsilon/3$ ($c_s^2=1/3$), $\delta_{\pi\pi} = 4\tau_\pi/3$, with $\eta/s$ and $\tau_\pi T$ being constants (here, $T \sim \varepsilon^{1/4}$ is the temperature and $s \sim T^3$ is the equilibrium entropy density). Assume, for simplicity, that
all the other transport coefficients vanish (as in \cite{Marrochio:2013wla}). The necessary conditions in \eqref{necessary_conditions} then impose that $\Lambda_a/(\varepsilon+P) \geq -1 + \frac{\eta}{s}\frac{1}{\tau_\pi T}$, so none of the eigenvalues of $\pi^\mu_\nu$ can be too negative. Also, 
when $\Lambda_a/(\varepsilon+P) > -1 + \frac{\eta}{s}\frac{1}{\tau_\pi T}$, the eigenvalues are also limited from above since \eqref{n1} gives $\Lambda_a/(\varepsilon+P)\leq 1- \frac{2}{\tau_\pi T}\frac{\eta}{s}$. Using typical values $\eta/s = 1/(4\pi)$ \cite{Kovtun:2004de} and $\tau_\pi T = 5\eta/s$ \cite{Denicol:2011fa} one then finds $-4/5 < \Lambda_a/(\varepsilon+P) \leq 3/5$. This implies that the relative magnitude of the shear stress tensor, $\sqrt{\pi_{\mu\nu}\pi^{\mu\nu}/(\varepsilon+P)^2}$, cannot be arbitrarily large. Therefore, heavy-ion simulations initiated with a NS Ansatz at an initial time $\tau_0$ where this relative magnitude $\sim (\eta/s)/(\tau_0 T(\tau_0,\vec{x}))$ necessarily violate causality in the regions where $1/(\tau_0 T(\tau_0,\vec{x}))$ is sufficiently large. This gives an example of the severe limitations on the far from equilibrium behavior of IS-like theories imposed by our novel nonlinear analysis.

\vskip 0.1cm
\textbf{5. Conclusions.}
In this work, we established for the first time that causality in fact holds for the full
set of nonlinear equations in IS-like theories without the need for symmetry assumptions and in the presence of both shear and bulk viscosity. All our conditions are simple algebraic inequalities among the dynamical variables that can be easily checked
in a given system or simulation.
Previous attempts to go beyond the linear regime were restricted to $1+1$ dimensions \cite{Denicol:2008ha} or assumed strong symmetry conditions \cite{Pu:2009fj,Floerchinger:2017cii}.
Without such restrictions, the only other work where nonlinear causality has been showed for IS-like systems is \cite{BemficaDisconziNoronha_IS_bulk}. The latter, however, included only bulk
viscosity and, thus, it is more important for applications in cosmology or neutron star mergers than in  
heavy-ion collisions. We have also studied the Cauchy problem for \eqref{conservaEM}-\eqref{supplemental},
establishing that it is well-defined, so that it is meaningful to talk about solutions. 

Prior to our work, one could only identify whether a numerical simulation of \eqref{conservaEM}-\eqref{supplemental}
violated causality if this caused (a) a breakdown of the simulation, (b) a manifestly spurious solution, or (c) clear non-physical behavior. These constraints are all too weak, as we now explain. 
For illustration, consider
$
- \partial_t^2\psi + (1+\psi) \Delta \psi = 0,
$
where $\Delta$ is the Laplacian.
This is a nonlinear wave equation with (nonlinear) speed given by $\sqrt{1+\psi}$ 
for \footnote{For $\psi < -1$, the equation is no longer a wave equation, becoming elliptic, and it is a degenerate
wave equation when $\psi = -1$.} $\psi > -1$. Indeed, the characteristics are given by 
$\xi_0 = \pm \sqrt{1+\psi} \,|\vec{\xi}|$. Therefore, the solutions are not causal 
when $\psi > 0$, but are causal for $-1 < \psi \leq 0$.
Nevertheless, the equation remains hyperbolic as long as $\psi > -1$.
Standard hyperbolic theory (see, e.g., \cite{SoggeNLWBook}) ensures that, given smooth initial 
data $\left. \psi \right|_{t=0}$ and $\left. \partial_t \psi \right|_{t=0}$, there exists a unique
smooth solution defined for some time. So any numerical scheme that is able to track
the unique solution will produce results in both the acausal and causal cases
$\psi > 0$ and $-1 < \psi \leq 0$, respectively. This makes it extremely difficult to 
infer violations of causality using (a) or (b) as criteria.
Exactly the same situation can happen in simulations of \eqref{conservaEM}-\eqref{supplemental}.
We also note that  linearizing the equation about the ``equilibrium" $\psi=0$ gives
$-\delta \psi_{tt} + \Delta \delta \psi = 0$, which is always causal, reinforcing again the idea
that causality cannot always be obtained from linearizations.

Criteria (c) also has limited applicability. First, there are different mechanisms that can produce non-physical solutions. Thus, it is still important to understand if unphysical behavior is being caused by causality violation, or some other mechanism, such as running beyond the limit where the effective description is valid. 
Second, relativistic fluids in the far from equilibrium regime, such as the QGP, may exhibit  
unexpected behavior, so one needs to be careful to differentiate genuine 
exotic features from those that are consequences of running a simulation in a superluminal regime. This may be particularly relevant to heavy-ion simulations where the values of the fields drop extremely rapidly at the edges of the QGP at early times and in the cold/dilute regions of plasma where a rescaling of dissipative tensors has been employed \cite{Schenke:2011bn,Bozek:2011ua,Shen:2014vra,Bazow:2016yra}.
Third, numerical simulations of relativistic fluids must be based on equations of motion that respect  causality, a fundamental physical principle in relativity. 

The results we presented here address all these difficulties, as one can
check if (A1)-(A3), \eqref{necessary_conditions}, or \eqref{conditions} hold
at any moment in numerical simulations \footnote{Comparing with the example of the equation
for $\psi$ above, this would be similar to monitor the value of $\sqrt{1+\psi}$:
if $\psi>1$, then the system is not causal, which is the analogue of
\eqref{necessary_conditions}, whereas causality is guaranteed if $-1< \psi \leq  0$, 
which is the analogue of \eqref{conditions}.} since all the quantities involved
in our inequalities can be readily extracted in numerical simulations \cite{Romatschke:2017ejr}.
We also note that our results apply, in particular, to the initial conditions, so 
\eqref{necessary_conditions} and \eqref{conditions} can be used 
to rule out initial conditions that violate causality or to select initial conditions for which causality
holds. This can be particularly relevant to further constrain the physical assumptions behind the modeling of initial conditions in QGP simulations.

In sum, in this Letter we established, for the first time in the literature, conditions to settle the longstanding questions concerning causality in Israel-Stewart-theories in the nonlinear, far-from-equilibrium regime. As such, our general results provide the most stringent tests to date to determine the validity of relativistic fluid dynamic approaches in heavy-ion collisions, astrophysics, and cosmology.


\vskip 0.1cm
\textbf{Acknowledgements.} MMD is partially supported by a Sloan Research Fellowship provided by the Alfred P. Sloan foundation, NSF grant DMS-1812826, and a Discovery Grant administered by Vanderbilt University. 
VH's work on this project was funded (full or in-part) by the University of Texas at San Antonio, Office of the Vice President for Research, Economic Development, and Knowledge Enterprise.  VH acknowledges partial support by NSF grants DMS-1614797 and DMS-1810687.

\appendix

\section*{Supplemental material}

In this Supplemental Material, in Section II 
we provide the proof that conditions (4) 
are necessary for causality, in Section III 
we provide the proof that conditions (5)
are sufficient for causality, and in Section IV 
we establish local existence
and uniqueness of solutions to the initial-value problem for equations (1)-(2). All these results
depend on a careful analysis of the roots of the characteristic equation
$\det(A^ \alpha\xi_\alpha)=0$. Thus, we first present in Section I 
a suitable factorization of $\det(A^ \alpha\xi_\alpha)$. 
In Section V 
we show that conditions (4), albeit necessary, are not sufficient for causality.
In Section VI 
we provide the formal definition of causality and
comment on why, in our case, it can be reduced to conditions (C1) and (C2).
Since causality is intrinsically tied to concepts of relativity theory,
we refer to the standard literature (e.g., \cite{HawkingEllisBook}) for further background.
Throughout this Supplemental Material, we continue to use the notation and definitions of the paper.

\section{I. The characteristic equation\label{S:Characteristic}}
Define 
$b=u^\alpha\xi_\alpha$, $v^\mu=\Delta^{\mu\nu}\xi_\nu$,  and $w^\mu=\pi^{\mu\nu}\xi_\nu$.
In terms of these quantities, the characteristic determinant can be written as
\bea
\label{detA}
&&\det(A^\alpha \xi_\alpha)=b^{13}\tau_\pi^{16}\tau_\Pi\det\bbm
b & \rho\xi_\nu+w_\nu \\
bc_s^2v^{\mu} & \rho b^2 \delta^\mu_\nu -b w_\nu u^\mu-\frac{\bar{C}^{\mu}_\nu}{\tau_\pi}-\frac{v^\mu\tilde{E}_\nu}{\tau_\Pi} 
\ebm=b^{14}\tau_\pi^{16}\tau_\Pi\det\left [M \right ],
\eea
where $M=[M^\mu_\nu]_{4\times 4}$ with $M^\mu_\nu=\rho b^2 \delta^\mu_\nu -b w_\nu u^\mu-\frac{\bar{C}^{\mu}_\nu}{\tau_\pi}-\frac{v^\mu\tilde{E}_\nu}{\tau_\Pi}-c_s^2v^{\mu}(\rho\xi_\nu+w_\nu)$, $\tilde{E}_\nu=E^\alpha_\nu\xi_\alpha=\left (\zeta+\delta_{\Pi\Pi}\Pi\right )\xi_\nu+\lambda_{\Pi \pi} w_\nu$, and 
\bea
\bar{C}^{\delta}_\nu&=&C^{\sigma\delta\alpha}_\nu\xi_\alpha\xi_\sigma=\frac{1}{2}\left [(2 \eta+\lambda_{\pi\Pi}\Pi)\xi_\mu\delta^\delta_\lambda+\frac{\tau_{\pi\pi}}{2}w_\lambda\delta^{\delta}_\mu+\frac{\tau_{\pi\pi}}{2}\pi_\lambda^{\delta}\xi_\mu\right ]\left (v^{\mu}\delta^\lambda_\nu+v^{\lambda}\delta^\mu_\nu-\frac{2}{3}\Delta^{\mu\lambda}\xi_\nu\right )\nonumber\\
&&-\frac{\tau_{\pi\pi}}{3}v^\delta w_\nu+\delta_{\pi \pi} w^\delta \xi_\nu-b\tau_\pi(w_\nu u^\delta+b\pi^\delta_\nu ).
\eea

Since $\pi^{\mu\nu}$ is symmetric and traceless, it can be diagonalized at any point in spacetime. The eigenvalue problem $\pi^\mu_\nu e^\nu_A=\Lambda_A e^\mu_A$, with $A=0,1,2,3$, defines an orthonormal set of eigenvectors $e_{A=0}^\mu=u^\mu$, $e^\mu_{A=a}=e^\mu_{a}$ with real eigenvalues $\Lambda_a$ for $a=1,2,3$ in the sense that $g_{\mu\nu}e_A^\mu e_B^\nu=\eta_{AB}$ where $\eta_{AB}=diag(-1,1,1,1)$. The eigenvalues are such that $\Lambda_0=0$ and $\Lambda_1+\Lambda_2+\Lambda_3=0$. Without any loss of generality, let us take $\Lambda_1\le \Lambda_2\le \Lambda_3$ with $\Lambda_1\le 0 \le \Lambda_3$ so that the trace is kept zero (note that if $\pi^\mu_\nu\neq 0$, this allows degeneracies to occur with multiplicity up to two). Since $\{ e_A^\mu \}$ is a complete set in $\mathbb{R}^4$, we may define a tetrad of dual vectors $\{ e_\nu^A \}$ by setting $ e^A_\nu \equiv\eta^{AB} (e_B)_\nu$ so that~\footnote{From now on, repeated Latin indexes are not summed unless explicitly stated.} $\delta_A^B = e_A^\nu e^B_\nu$. Also, the following completeness relation holds: $\delta^{\mu}_\nu = \sum_A e_A^\mu e^A_\nu  = -u^\mu u_\nu+\sum_a e_a^\mu (e_a)_\nu$. Therefore, the components of any four-vector $z^\mu$ relative to the tetrad $\{ e_A^\mu \}$ are defined by $z^A \equiv z^\nu e_\nu^A$. We can then use this to define $v_A \equiv e_A^\mu v_\mu$ and $\xi_A\equiv e^\mu_A\xi_\mu$. Given that $\xi^\mu=-b u^\mu+\sum_a v^a e_a^\mu$ ($a=1,2,3$) one finds that $\xi_{A=0}=-\xi^{A=0}=b$ while $\xi_a=v_a$. Furthermore, $w_A \equiv e_A^\mu w_\mu=e_A^\mu\pi_{\mu\nu}\xi^\nu=\Lambda_A \xi_A = \Lambda_A v_A$, where we used that $\Lambda_0=0$ and again $\xi_a = v_a$ (note also that $v^a=v_a$ since $\eta_{ab}=\delta_{ab}$). Using these observations,
we can show that the determinant $\det(M)$ needed for the characteristics in \eqref{detA} is given by 
\bea
\label{1}
&&\det(M)=\det(E^{-1}ME)=m_0 m_1 m_2 m_3\nonumber \\ &\times& \Bigg[1-\sum_a\frac{\left \{\frac{1}{6\tau_\pi}[2 \eta+\lambda_{\pi\Pi}\Pi+(6\delta_{\pi\pi}-\tau_{\pi\pi})\Lambda_a]+\frac{\zeta+\delta_{\Pi\Pi}\Pi+\lambda_{\Pi \pi}\Lambda_a}{\tau_\Pi}+(\rho+\Lambda_a)c_s^2\right \}\hat{v}_a^2}{\bar{m}_a}\nonumber\\
&&-\frac{12\delta_{\pi\pi}-\tau_{\pi\pi}}{12\tau_\pi}\left (\frac{\lambda_{\Pi \pi} }{\tau_\Pi}+c_s^2-\frac{\tau_{\pi\pi}}{12\tau_\pi}\right )\sum_{\underset{a < b}{a,b}}\frac{(\Lambda_a-\Lambda_b)^2\hat{v}_a^2\hat{v}_b^2}{\bar{m}_a\bar{m}_b}\Bigg],
\eea
where $E=[e^\mu_A]_{4\times 4}$, $E^{-1}=[e^B_\nu]_{4\times4}$, and $E^{-1}ME=[e^A_\mu M^\mu_\nu e^\nu_B]_{4\times 4}$. Also, we defined above $m_0=\rho \left (b^2 -\sum_{a}\mathfrak{g}_{a}v_a^2\right )$, $\mathfrak{g}_{a}=\frac{2(2 \eta+\lambda_{\pi\Pi}\Pi)+\tau_{\pi\pi}\Lambda_a}{4 \rho \tau_\pi}$, $m_a=(\rho+\Lambda_a) b^2-\frac{1}{2\tau_\pi}(2 \eta+\lambda_{\pi\Pi}\Pi)(v\cdot v)-\frac{\tau_{\pi\pi}}{4\tau_\pi}\left (\Lambda_a v\cdot v+\sum_c \Lambda_c v_c^2\right )$, $\hat{v}_a=v_a/\sqrt{v\cdot v}$ (assuming $v\neq 0$), and $\bar{m}_0 = m_0/(v \cdot v), \bar{m}_a=m_a/(v\cdot v)$. Note that $\sum_a\hat{v}_a^2=\sum_a v_a^2/(v\cdot v)=1$ since $v\cdot v=v^\mu v_\mu=\sum_a v_a^2$. Assuming $v\neq 0$ is allowed because $v=0$ does not lead to nontrivial roots $b\neq 0$ of the characteristic equation if assumptions (A1)--(A3) hold. 

The roots $\xi$ of $\det(A^ \alpha\xi_\alpha)=0$ defined in Eq. \eqref{detA} are the fourteen roots coming from $b=u^ \alpha\xi_\alpha=0$ together with 8 roots from $\det(M)=0$ in Eq. \eqref{1} which consist of the 2 roots from $m_0=0$ and the 6 roots coming from the zeros of 
\be
\label{4-1}
f(k)=\bar{m}_1\bar{m}_2\bar{m}_3 G(k),
\ee
where we defined $k\equiv b^2/v\cdot v$ and
\bea
\label{4-2}
&&G(k)=1-\sum_a\frac{\left \{\frac{1}{6\tau_\pi}[2 \eta+\lambda_{\pi\Pi}\Pi+(6\delta_{\pi\pi}-\tau_{\pi\pi})\Lambda_a]+\frac{\zeta+\delta_{\Pi\Pi}\Pi+\lambda_{\Pi \pi}\Lambda_a}{\tau_\Pi}+(\rho+\Lambda_a)c_s^2\right \}\hat{v}_a^2}{\bar{m}_a}\nonumber\\
&&-\frac{12\delta_{\pi\pi}-\tau_{\pi\pi}}{12\tau_\pi}\left (\frac{\lambda_{\Pi \pi} }{\tau_\Pi}+c_s^2-\frac{\tau_{\pi\pi}}{12\tau_\pi}\right )\sum_{\underset{a < b}{a,b}}\frac{(\Lambda_a-\Lambda_b)^2\hat{v}_a^2\hat{v}_b^2}{\bar{m}_a\bar{m}_b}.
\eea  
In this notation $\det(M)=m_0 (v\cdot v)^3 f(k)$ because we used the definition $\bar{m}_a=m_a/v\cdot v$. Note that although $G(k)$ has $\bar{m}_a$ appearing in denominators, these are canceled by the multiplication of $G(k)$ by $\bar{m}_1 \bar{m}_2 \bar{m}_3$ in the definition of $f(k)$. Thus, $f(k)$ is a polynomial of degree
3 in $k$ (of degree 6 in $b$) and is defined for all values of $k\in \mathbb{R}$. Then, it is possible to factorize $f(k)$ as
\be
\label{4}
f(k)=\left [\prod_{a=1}^3(\varepsilon+P+\Pi+\Lambda_a)\right ](k-k_1)(k-k_2)(k-k_3),
\ee
where $k_1, k_2, k_3$ as the three roots of $f(k)$. Note that for the sake of brevity, we have suppressed the dependence on $\hat v$ in writing $G(k)$ and $f(k)$ (to be more precise, these should have been written as $G(k, \hat v), f(k, \hat v)$).

Conditions (C1) and (C2) for causality demand that all the 22 roots
$\xi_0 = \xi_0(\xi_i)$ of $\det(A^ \alpha\xi_\alpha)=0$ 
are real and satisfy $\xi_\alpha\xi^\alpha=-b^2+v\cdot v\ge0$, i.e., $0\le k\le 1$. The 14 roots $b=0$ are causal. Thus, the rest the analysis of necessary conditions
in Section II 
will focus on the remaining roots defined by $f(k) = 0$. We summarize this in the following important statement:
\be
\begin{array}{l}
\text{The system is causal if and only if for all for all $\hat v$ on the unit sphere, the roots}\\
\text{of $\bar{m}_0(k, \hat v) = 0$ and $f(k, \hat v)=0$ are real and $0\leq k \leq 1$.}
\end{array}\tag{C3}  \label{Condition:C3}
\ee

\section{II. Derivation of necessary conditions for causality\label{S:Necessary}}

Here we establish that conditions (4) are necessary (but not sufficient, see Section V) 
for causality. More precisely, we establish the following Theorem.

\begin{theorem}
\label{T:Necessary}
Let $\Psi=(\varepsilon,u^\nu,\Pi,\pi^{0\nu},\pi^{1\nu},\pi^{2\nu},\pi^{3\nu})_{\nu=0,\dots,3}$
be a smooth solution to 
equations (1)-(2) in Minkowski space, with $u_\mu u^\mu = -1$ and 
$\pi_{\mu\nu}$ satisfying $\pi^\mu_\mu = 0$ and $u^\mu \pi_{\mu\nu} = 0$. 
Suppose that (A1)-(A3) hold. If any of conditions (4) is not satisfied, then
$\Psi$ is not causal in the sense of Definition \ref{D:Causality} (see Section VI).
\end{theorem}

\noindent \emph{Proof of Theorem \ref{T:Necessary}:}
Our derivation of necessary conditions for causality is via the following reasoning. Causality requires
that conditions (C1) and (C2) hold for all $\xi_i$. Thus, in order to violate causality, it suffices to show
that for some $\xi_i$, (C1) or (C2) fails. Suppose now that we find a condition, say $\mathsf{Z}$, for which
we can exhibit one $\xi_i$ such that (C1) or (C2) fail, i.e., we obtain the statement 
``$\mathsf{Z}$ implies non-causality." This statement is logically equivalent to ``Causality implies 
non-$\mathsf{Z}$."
In other other, non-$\mathsf{Z}$ is a necessary condition for causality: if it is violated, the system is not
causal. In our case, conditions like $\mathsf{Z}$ will be inequalities among the scalars of the problem (e.g., the relaxation times, eigenvalues $\Lambda_a$, etc.) of the form $A > B$, whose negation 
is then $A \leq B$. The latter
is then the necessary condition we are looking for: if $A \leq B$ does not hold, the system is not causal.

Recall that (C1) and (C2) is equivalent to \eqref{Condition:C3}, so in view of the foregoing discussion, we aim to violate \eqref{Condition:C3}.
With the choice $\hat{v}_a=\delta_{ad}$, one can write $m_0=\rho (v\cdot v)(k-\mathfrak{g}_d)=0$.
Under our assumptions, the only root is $k = \mathfrak{g}_d$.
Since we need $0 \leq k \leq 1$, as discussed, and since $\mathfrak{g}_1 \leq \mathfrak{g}_2 \leq
\mathfrak{g}_3$, 
causality if violated if $\mathfrak{g}_1 < 0$, leading to condition (4a), or if 
$\mathfrak{g}_3 > 1$, leading to condition (4b). Observe also that if $\rho$ were allowed to vanish, then
the characteristic determinant would also vanish, leading to non-causality. See our discussion
of the condition $\varepsilon+P+\Pi > 0$ in the main text.

As for the roots of $f(k)$, we may note that now in $f(k)=\bar{m}_1\bar{m}_2\bar{m}_3 G(k)$ 
we have
\be
\bar{m}_a=(\varepsilon+P+\Pi+\Lambda_a) k-\frac{1}{2\tau_\pi}(2 \eta+\lambda_{\pi\Pi}\Pi)-\frac{\tau_{\pi\pi}}{4\tau_\pi}\left (\Lambda_a +\Lambda_d\right )
\ee
and
\bea
&&G(k)=1-\frac{\left \{\frac{1}{6\tau_\pi}[2 \eta+\lambda_{\pi\Pi}\Pi+(6\delta_{\pi\pi}-\tau_{\pi\pi})\Lambda_d]+\frac{\zeta+\delta_{\Pi\Pi}\Pi+\lambda_{\Pi \pi}\Lambda_d}{\tau_\Pi}+(\rho+\Lambda_d)c_s^2\right \}}{\bar{m}_d}
\eea
because we have set $\hat{v}_a=\delta_{ad}$. We may therefore rewrite 
\bea
&&f(k)=\bar{m}_a\bar{m}_b\bigg [\bar{m}_d-\bigg \{\frac{1}{6\tau_\pi}[2 \eta+\lambda_{\pi\Pi}\Pi+(6\delta_{\pi\pi}-\tau_{\pi\pi})\Lambda_d]+\frac{\zeta+\delta_{\Pi\Pi}\Pi+\lambda_{\Pi \pi}\Lambda_d}{\tau_\Pi}\nonumber\\
&&+(\rho+\Lambda_d)c_s^2\bigg \}\bigg ],\label{nf}
\eea
where $a\ne b$ and $a,b\ne d$. Setting each of the factors $m_a, m_b$ equal to zero, we obtain the roots
\be
\label{ma}
 k=\frac{\frac{1}{2\tau_\pi}(2 \eta+\lambda_{\pi\Pi}\Pi)+\frac{\tau_{\pi\pi}}{4\tau_\pi}\left (\Lambda_a +\Lambda_d\right )}{\varepsilon+P+\Pi+\Lambda_a},\quad a\ne d.
 \ee
Causality is violated if $k< 0$, leading to condition (4c), of if $k>1$, leading to condition (4d).
The remaining root in \eqref{nf} is obtained when the term in brackets vanishes, giving
\bea
\label{nf2}
&& k=\frac{\frac{1}{2\tau_\pi}(2 \eta+\lambda_{\pi\Pi}\Pi)+\frac{\tau_{\pi\pi}}{2\tau_\pi}\Lambda_d}{\varepsilon+P+\Pi+\Lambda_d}\nonumber\\
&&+\frac{\bigg \{\frac{1}{6\tau_\pi}[2 \eta+\lambda_{\pi\Pi}\Pi+(6\delta_{\pi\pi}-\tau_{\pi\pi})\Lambda_d]+\frac{\zeta+\delta_{\Pi\Pi}\Pi+\lambda_{\Pi \pi}\Lambda_d}{\tau_\Pi}+(\rho+\Lambda_d)c_s^2\bigg \}}{\varepsilon+P+\Pi+\Lambda_d}.
\eea
Causality is violated if $k< 0$, leading to (4e), or if $k> 1$, leading to (4f). This finishes
the proof. \hfill \qed

\bigskip

We remark that the diagonalization of $\pi_{\mu\nu}$ was carried out in terms of 
orthonormal frames which can be defined for any Lorentzian metric. 
Also, our computations are manifestly covariant. 
Thus, the result of Theorem  \ref{T:Necessary}
remains true in a general globally hyperbolic space-time, as mentioned in the main text.
This includes, in particular, the cases where the equations hold in a globally hyperbolic
subset of Minkowski space or in
 $I\times \mathbb{T}^3$ with the Minkowski metric,
where $I \subseteq \mathbb{R}$ is an interval and $\mathbb{T}^3$ is the three-dimensional torus.

\section{III. Derivation of sufficient conditions for causality\label{S:Sufficient}}

Here we establish that conditions (5) are sufficient for causality. 
More precisely,
we establish the following Theorem.

\begin{theorem}
\label{T:Sufficient}
Let $\Psi=(\varepsilon,u^\nu,\Pi,\pi^{0\nu},\pi^{1\nu},\pi^{2\nu},\pi^{3\nu})_{\nu=0,\dots,3}$
be a smooth solution to 
equations (1)-(2) in Minkowski space,  with $u_\mu u^\mu = -1$ and 
$\pi_{\mu\nu}$ satisfying $\pi^\mu_\mu = 0$ and $u^\mu \pi_{\mu\nu} = 0$. 
Suppose that (A1)-(A3) and (5) hold. Then
$\Psi$ is causal in the sense of Definition \ref{D:Causality}.
\end{theorem}

\noindent \emph{Proof of Theorem \ref{T:Sufficient}:} As discussed in Section I, 
the 14 roots $b=0$ are causal and do not need any further treatment. The remaining 8 roots 
that come from $\det(M)=0$ are, again, the two roots of $m_0$  and the six roots of $f(k)$ defined in \eqref{4-1}. We begin by analyzing the two roots of $m_0$. Recalling  that $v=0$ does not lead a nontrivial root  
of $\det(A^ \alpha\xi_\alpha)=0$, we see that the roots of $m_0$ are given by $b^2 = k =\sum_a\mathfrak{g}_a\hat{v}^2_a$. 
For these roots we need to check (according to \eqref{Condition:C3}) that 
\be
\label{1-1}
0\le \sum_{a}\mathfrak{g}_{a}\hat{v}^2_a\le 1.
\ee 
(A3) together with conditions (5a) and (5b) give $0\leq \mathfrak{g}_{1}\le \mathfrak{g}_{2}\le\mathfrak{g}_{3}\le 1$. From $\mathfrak{g}_1\le \sum_{a} \mathfrak{g}_a\hat{v}_a^2\le\mathfrak{g}_3$, we see that \eqref{1-1} is satisfied.

Now we analyze the remaining 6 roots of $\det(M)=0$ coming from $f(k)$ defined in Eq.~\eqref{4-1} and written explicitly as a polynomial in \eqref{4}. We will show further below that
the three roots $k_i$ in \eqref{4} are real. But let us first show that any real 
root of $f$ must lie within $[0,1]$.
Since $f$ is a cubic polynomial, it either has only one real root, say $s_1$, or three
real roots, in which case we can order them as $k_1\leq k_2 \leq k_3$ in \eqref{4}.
Invoking  (5a), we see that in the first case $f$ is negative to the left of $s_1$ and positive to its
right, and in the second case 
that $f$ is a growing
cubic polynomial except in the interval between the roots $k_1$ and $k_3$.
In either situation, any real root will be between $0$ and $1$ if
\be
\label{5-1}
f(k<0)< 0,
\ee
and
\be
\label{5-2}
f(k>1)> 0.
\ee
Let us first verify the inequality \eqref{5-2}. For $k>1$
\bea
&&\bar{m}_a(k>1) \ge k(\varepsilon+P+\Pi-|\Lambda_1|)-\frac{1}{2\tau_\pi}(2 \eta+\lambda_{\pi\Pi}\Pi)-\frac{\tau_{\pi\pi}}{2\tau_\pi}\Lambda_3
\eea
where we have used $-2|\Lambda_1|\le \Lambda_a +\sum_c \Lambda_c \hat{v}_c^2\le 2\Lambda_3$. Now, observe that
\[k(\varepsilon+P+\Pi-|\Lambda_1|)-\frac{1}{2\tau_\pi}(2 \eta+\lambda_{\pi\Pi}\Pi)-\frac{\tau_{\pi\pi}}{2\tau_\pi}\Lambda_3> (\varepsilon+P+\Pi-|\Lambda_1|)-\frac{1}{2\tau_\pi}(2 \eta+\lambda_{\pi\Pi}\Pi)-\frac{\tau_{\pi\pi}}{2\tau_\pi}\Lambda_3\]
for $k>1$, hence the condition (5a) lead us to $\bar{m}_a(k\ge1)>0$. 
This guarantees that 
\begin{align}
\bar{m}_1(k>1)\bar{m}_2(k>1)\bar{m}_3(k>1)>0.
\nonumber
\end{align}
To obtain $f(k>1)>0$ in \eqref{5-2}, we therefore need $G(k>1)>0$. By means of (5c) and (5d),
\bea
&&-\sum_a\frac{\left \{\frac{1}{6\tau_\pi}[2 \eta+\lambda_{\pi\Pi}\Pi+(6\delta_{\pi\pi}-\tau_{\pi\pi})\Lambda_a]+\frac{\zeta+\delta_{\Pi\Pi}\Pi+\lambda_{\Pi \pi}\Lambda_a}{\tau_\Pi}+(\rho+\Lambda_a)c_s^2\right \}\hat{v}_a^2}{\bar{m}_a(k>1)}
\nonumber\\
&&> -\frac{\frac{1}{6\tau_\pi}[2 \eta+\lambda_{\pi\Pi}\Pi+(6\delta_{\pi\pi}-\tau_{\pi\pi})\Lambda_3]+\frac{\zeta+\delta_{\Pi\Pi}\Pi+\lambda_{\Pi \pi}\Lambda_3}{\tau_\Pi}+(\varepsilon+P+\Pi+\Lambda_3)c_s^2 }{\varepsilon+P+\Pi-|\Lambda_1|-\frac{1}{2\tau_\pi}(2 \eta+\lambda_{\pi\Pi}\Pi)-\frac{\tau_{\pi\pi}}{2\tau_\pi}\Lambda_3}
\eea
as well as
\bea
&&-\frac{12\delta_{\pi\pi}-\tau_{\pi\pi}}{12\tau_\pi}\left (\frac{\lambda_{\Pi \pi} }{\tau_\Pi}+c_s^2-\frac{\tau_{\pi\pi}}{12\tau_\pi}\right )\sum_{a<b}\frac{(\Lambda_a-\Lambda_b)^2\hat{v}_a^2\hat{v}_b^2}{\bar{m}_a(k>1)\bar{m}_b(k>1)}\nonumber\\
&&>-\frac{\frac{12\delta_{\pi\pi}-\tau_{\pi\pi}}{12\tau_\pi}\left (\frac{\lambda_{\Pi \pi} }{\tau_\Pi}+c_s^2-\frac{\tau_{\pi\pi}}{12\tau_\pi}\right )(\Lambda_3-\Lambda_1)^2}{\left [\varepsilon+P+\Pi-|\Lambda_1|-\frac{1}{2\tau_\pi}(2 \eta+\lambda_{\pi\Pi}\Pi)-\frac{\tau_{\pi\pi}}{2\tau_\pi}\Lambda_3\right ]^2},
\eea
and thus,
\bea
&&G(k>1)> 1-\frac{\frac{1}{6\tau_\pi}[2 \eta+\lambda_{\pi\Pi}\Pi+(6\delta_{\pi\pi}-\tau_{\pi\pi})\Lambda_3]+\frac{\zeta+\delta_{\Pi\Pi}\Pi+\lambda_{\Pi \pi}\Lambda_3}{\tau_\Pi}+(\varepsilon+P+\Pi+\Lambda_3)c_s^2 }{\varepsilon+P+\Pi-|\Lambda_1|-\frac{1}{2\tau_\pi}(2 \eta+\lambda_{\pi\Pi}\Pi)-\frac{\tau_{\pi\pi}}{2\tau_\pi}\Lambda_3}\nonumber\\
&&-\frac{\frac{12\delta_{\pi\pi}-\tau_{\pi\pi}}{12\tau_\pi}\left (\frac{\lambda_{\Pi \pi} }{\tau_\Pi}+c_s^2-\frac{\tau_{\pi\pi}}{12\tau_\pi}\right )(\Lambda_3+|\Lambda_1|)^2}{\left [\varepsilon+P+\Pi-|\Lambda_1|-\frac{1}{2\tau_\pi}(2 \eta+\lambda_{\pi\Pi}\Pi)-\frac{\tau_{\pi\pi}}{2\tau_\pi}\Lambda_3\right ]^2}.
\eea
Note that we have used $\max_{a,b} (\Lambda_a - \Lambda_b)^2=(\Lambda_3-\Lambda_1)^2 = (\Lambda_3 + |\Lambda_1|)^2$, which follows from the ordering of the eigenvalues $\Lambda_a$. Hence (5e) implies $G(k)> 0$ for $k > 1$.

It now remains to verify the inequality \eqref{5-1}. In this case, when $k<0$
\bea
\label{6}
&&\bar{m}_a(k<0)=-|k|(\varepsilon+P+\Pi+\Lambda_a)-\frac{1}{2\tau_\pi}(2 \eta+\lambda_{\pi\Pi}\Pi)-\frac{\tau_{\pi\pi}}{4\tau_\pi}\left (\Lambda_a +\sum_c \Lambda_c \hat{v}_c^2\right )\nonumber\\
&&<-\frac{1}{2\tau_\pi}(2 \eta+\lambda_{\pi\Pi}\Pi)+\frac{\tau_{\pi\pi}}{2\tau_\pi}|\Lambda_1|.
\eea
From condition (5b), one has that $\bar{m}_a(k\le0)<0$. Then, \[f(k<0)=\bar{m}_1(k<0)\bar{m}_2(k<0)\bar{m}_3(k<0)G(k<0)< 0\] if, and only if, $G(k<0)> 0$. 
Due to $\bar{m}_a(k\le0)<0$ together with (5c) and (5d), we obtain that
\bea
&&\sum_a\frac{\left \{\frac{1}{6\tau_\pi}[2 \eta+\lambda_{\pi\Pi}\Pi+(6\delta_{\pi\pi}-\tau_{\pi\pi})\Lambda_a]+\frac{\zeta+\delta_{\Pi\Pi}\Pi+\lambda_{\Pi \pi}\Lambda_a}{\tau_\Pi}+(\rho+\Lambda_a)c_s^2\right \}\hat{v}_a^2}{-\bar{m}_a(k<0)}
\nonumber\\
&&> \frac{\frac{1}{6\tau_\pi}[2 \eta+\lambda_{\pi\Pi}\Pi-(6\delta_{\pi\pi}-\tau_{\pi\pi})|\Lambda_1|]+\frac{\zeta+\delta_{\Pi\Pi}\Pi-\lambda_{\Pi \pi}|\Lambda_1|}{\tau_\Pi}+(\varepsilon+P+\Pi-|\Lambda_1|)c_s^2 }{-m_a(k<0)}.
\eea
Condition (5f) guarantees that $\sum_{a}\ldots > 0$ in the above inequality. Moreover, 
\bea
&&-\frac{12\delta_{\pi\pi}-\tau_{\pi\pi}}{12\tau_\pi}\left (\frac{\lambda_{\Pi \pi} }{\tau_\Pi}+c_s^2-\frac{\tau_{\pi\pi}}{12\tau_\pi}\right )\sum_{a<b}\frac{(\Lambda_a-\Lambda_b)^2\hat{v}_a^2\hat{v}_b^2}{\bar{m}_a(k<0)\bar{m}_b(k<0)}\nonumber\\
&&>-\frac{\frac{12\delta_{\pi\pi}-\tau_{\pi\pi}}{12\tau_\pi}\left (\frac{\lambda_{\Pi \pi} }{\tau_\Pi}+c_s^2-\frac{\tau_{\pi\pi}}{12\tau_\pi}\right )(\Lambda_3+|\Lambda_1|)^2}{\left [\frac{1}{2\tau_\pi}(2 \eta+\lambda_{\pi\Pi}\Pi)-\frac{\tau_{\pi\pi}}{2\tau_\pi}|\Lambda_1|\right ]^2}.
\eea
where we used \eqref{6} and $(\Lambda_3+|\Lambda_1|)^2 = \max_{a,b} (\Lambda_a - \Lambda_b)^2$ again.
Now, since
\bea
&&G(k<0)> 1-\frac{\frac{12\delta_{\pi\pi}-\tau_{\pi\pi}}{12\tau_\pi}\left (\frac{\lambda_{\Pi \pi} }{\tau_\Pi}+c_s^2-\frac{\tau_{\pi\pi}}{12\tau_\pi}\right )(\Lambda_3+|\Lambda_1|)^2}{\left [\frac{1}{2\tau_\pi}(2 \eta+\lambda_{\pi\Pi}\Pi)-\frac{\tau_{\pi\pi}}{2\tau_\pi}|\Lambda_1|\right ]^2},
\eea
we have $G(k<0)> 0$ from condition (5g), finally implying $f(k<0)<0$. 

It remains to establish the reality of the roots $k_i$ in \eqref{4}.
To do that, let us write $G(k)$ as
\bea
\label{G}
G(k)&=&1-\sum_a\frac{R_a\hat{v}_a^2}{\bar{m}_a}-\sum_{\underset{a < b}{a,b}}\frac{S_{ab}\hat{v}_a^2\hat{v}_b^2}{\bar{m}_a\bar{m}_b}
\eea
and
\be
\bar{m}_a=\rho_ak-r_a,
\ee
where
\bea
R_a&=&\frac{1}{6\tau_\pi}[2 \eta+\lambda_{\pi\Pi}\Pi+(6\delta_{\pi\pi}-\tau_{\pi\pi})\Lambda_a]+\frac{\zeta+\delta_{\Pi\Pi}\Pi+\lambda_{\Pi \pi}\Lambda_a}{\tau_\Pi}+(\rho+\Lambda_a)c_s^2\\
S_{ab}&=&\frac{12\delta_{\pi\pi}-\tau_{\pi\pi}}{12\tau_\pi}\left (\frac{\lambda_{\Pi \pi} }{\tau_\Pi}+c_s^2-\frac{\tau_{\pi\pi}}{12\tau_\pi}\right )(\Lambda_a-\Lambda_b)^2,\\
\rho_a&=&\rho+\Lambda_a=\varepsilon+P+\Pi+\Lambda_a,\\
r_a&=&\frac{1}{2\tau_\pi}(2 \eta+\lambda_{\pi\Pi}\Pi)+\frac{\tau_{\pi\pi}}{4\tau_\pi}\left (\Lambda_a +\sum_c \Lambda_c \hat{v}_c^2\right ).
\eea
Note, in particular, that $\bar{r}_1\le r_a\le \bar{r}_3$, where $\bar{r}_{1,3}\equiv\frac{1}{2\tau_\pi}(2 \eta+\lambda_{\pi\Pi}\Pi)+\frac{\tau_{\pi\pi}\Lambda_{1,3}}{2\tau_\pi}>0$ from (5b).
By applying conditions (5) one has that $R_a,S_{ab},\rho_a,r_a\ge0$. Then, $f(k)$ can be written as
\bea
&&f(k)=\bar{m}_1\bar{m}_2\bar{m}_3-\bar{m}_1\bar{m}_2R_3\hat{v}^2_3-\bar{m}_2\bar{m}_3R_1\hat{v}^2_1-\bar{m}_3\bar{m}_1R_2\hat{v}^2_2-\bar{m}_1S_{23}\hat{v}^2_2\hat{v}^2_3-\bar{m}_2S_{13}\hat{v}^2_1\hat{v}^2_3\nonumber\\
&&-\bar{m}_3S_{12}\hat{v}^2_1\hat{v}^2_2\nonumber\\
&&=a_3k^3+a_2k^2+a_1k+a_0,
\eea
where
\bea
a_0&=&-\big (r_1r_2r_3+r_1r_2 R_3\hat{v}^2_3+r_2r_3 R_1\hat{v}^2_1+r_1r_3R_2\hat{v}^2_2-r_1S_{23}\hat{v}^2_2\hat{v}^2_3-r_2S_{13}\hat{v}^2_1\hat{v}^2_3\nonumber\\
&&-r_3S_{12}\hat{v}^2_1\hat{v}^2_2\big ),\\
a_1&=&\rho_1r_2r_3+\rho_2r_1r_3+\rho_3r_1r_2+(\rho_1r_2+\rho_2r_1) R_3\hat{v}^2_3+(\rho_2r_3+\rho_3r_2) R_1\hat{v}^2_1\nonumber\\
&&+(\rho_3r_1+\rho_1r_3)R_2\hat{v}^2_2-\rho_1S_{23}\hat{v}^2_2\hat{v}^2_3-\rho_2S_{13}\hat{v}^2_1\hat{v}^2_3-\rho_3S_{12}\hat{v}^2_1\hat{v}^2_2,\\
a_2&=&-(\rho_1\rho_2 r_3+\rho_1\rho_3 r_2+\rho_2\rho_3 r_1+\rho_1\rho_2 R_3\hat{v}^2_3+\rho_2\rho_3 R_1\hat{v}^2_1+\rho_1\rho_3R_2\hat{v}^2_2),\\
a_3&=&\rho_1\rho_2\rho_3.
\eea
In view of (5), we have $a_3>0$ and $a_2<0$. Since all coefficients of $f(k)$ are real, then at least one of the roots must be real, say $k=s_1\in\mathbb{R}$ is the real root. Then, we know that the other two roots $s_2$ and $s_3$ are real or complex conjugate, i.e., $s_3^*=s_2$. Let us assume that $s_2$ and $s_3$ can be imaginary and set $s_{2,3}=k_R\pm i k_I$, $k_I \neq 0$. By using Vieta's formula $s_1+s_2+s_3=-\frac{a_2}{a_3}=\frac{|a_2|}{a_3}>0$ we obtain that 
\be
\frac{|a_2|}{a_3}-1\le 2k_R=\frac{|a_2|}{a_3}-s_1\le \frac{|a_2|}{a_3}.
\ee
Thus, the following condition holds,
\be
\label{Vieta}
\frac{3\rho_1(\bar{r}_1+R_1)}{\rho_2\rho_3}-1< 2k_R<\frac{3\rho_3(\bar{r}_3+R_3)}{\rho_1\rho_2}
\ee
because the real root $s_1\in[0,1]$ when (5a)--(5g) apply, as we have already showed. 
Since we are assuming $s_{2,3}=k_R\pm ik_I$, where $k_I\ne0$, we have that
 $\bar{m}_a(s_{2,3})=\rho_a k_R-r_a\pm i k_I$ cannot be zero (unless $k_I=0$ and the roots are real). Consequently, from \eqref{G} we obtain that $f(s_{2,3})=0$ lead us to $G(s_{2,3})=0$, where $s_{2,3}$ must obey the above conditions implied by $f$ being a cubic polynomial, 
in particular the condition on $k_R$ in \eqref{Vieta}. Thus, let us split $G(s_{2,3})$ in \eqref{6} into $G_R(s_{2,3})+iG_I(s_{2,3})$, where $G_R(s_{2,3})=\Re [G(s_{2,3})]$ and $G_I(s_{2,3})=\Im [G(s_{2,3})]$. In particular, 
\bea
\label{numerator}
G_I(s_{2,3})=\pm k_I\sum_a\frac{\hat{v}^2_a}{|\bar{m}_a|^2}\left [\rho_a R_a+\sum_{\underset{b>a}{b}}\frac{[\rho_a(\rho_bk_R-\bar{r}_b)+\rho_b(\rho_ak_R-\bar{r}_a)]S_{ab}\hat{v}^2_b}{|\bar{m}_b|^2}\right ].
\eea 
To show that the roots are real, if suffices to have $G_I(s_{2,3})\ne 0$. We distinguish two cases.
If $S_{ab}=0$ then $G_I(s_{2,3})\ne 0$ because we assumed $k_I\ne0$. This means that in this case the roots must all be real. On the other hand, if $S_{ab}\neq 0$ and $\rho_1R_1-\bar{r}_3>0$, then Eq.\ \eqref{numerator} also gives $G_I(s_{2,3})\ne 0$, because then the sum over $b$ in \eqref{numerator} is $>0$. To check that $\rho_1R_1-\bar{r}_3>0$, note first that (5a) guarantees that $\rho_a>r_a$. Then, by means of \eqref{Vieta}, we obtain that
\bea
\label{74}
\rho_1 k_R-\bar{r}_3&>&\frac{\rho_1}{2}\left (\frac{3\rho_1(R_1+\bar{r}_1)}{\rho_2\rho_3}-1-\frac{2\bar{r}_3}{\rho_1}\right )\ge0 
\eea 
because of condition (5h), and this implies $\rho_1k_R-\bar{r}_3>0$. Since we have already showed
that any real root of $f(k)$ must lie within $[0,1]$, this finishes our proof. \hfill\qed

\bigskip

We remark that the diagonalization of $\pi_{\mu\nu}$ was carried out in terms of 
orthonormal frames which can be defined for any Lorentzian metric. 
Also, our computations are manifestly covariant. 
Thus, the result of Theorem  \ref{T:Sufficient}
remains true in a general globally hyperbolic space-time, as mentioned in the main text.
This includes, in particular, the cases where the equations hold in a globally hyperbolic
subset of Minkowski space or in
 $I\times \mathbb{T}^3$ with the Minkowski metric,
where $I \subseteq \mathbb{R}$ is an interval and $\mathbb{T}^3$ is the three-dimensional torus.

\section{IV. Local existence and uniqueness\label{S:Well-posedness}}

In this Section, we establish the local existence and uniqueness of solutions to the 
Cauchy problem. Below, $\mathcal{G}$ is the space of Gevrey functions or 
quasi-analytic functions.

\begin{theorem}
\label{T:Existence}
Consider the Cauchy problem for equations (1)-(2) in 
Minkowski space,
with initial data
$\mathring{\Psi}=(\mathring{\varepsilon},\mathring{u}^\nu,\mathring{\Pi},\mathring{\pi}^{0\nu},
\mathring{\pi}^{1\nu},\mathring{\pi}^{2\nu},\mathring{\pi}^{3\nu})_{\nu=0,\dots,3}$
given on $\{ t = 0 \}$. Assume that the data satisfies the 
constraints\footnote{Alternatively, we could have only unconstrained data be prescribed and obtain
the full set of data from the stated constraints. For example, we could have $\mathring{u}^i$ prescribed
and define $u^0$ so that $\mathring{u}^\nu$ is unit time-like and future pointing.}
$\mathring{u}^\nu \mathring{u}_\nu = -1$, $\mathring{u}^\nu$ is future-pointing, 
$\mathring{\pi}^\nu_\nu = 0$,
and $\mathring{\pi}^\nu_\mu \mathring{u}^\mu = 0$. Suppose that (A1)-(A3) and (5) hold for
$\mathring{\Psi}$ in a strict form (i.e. $<$ instead of $\le$, $>$ instead of $\ge$). Finally, assume that $\mathring{\Psi} \in \mathcal{G}^{\delta}(\{ t=0\})$,
where $1 \leq \delta < 20/19$. Then, there exist
a $T>0$ and a unique 
$\Psi=(\varepsilon,u^\nu,\Pi,\pi^{0\nu},\pi^{1\nu},\pi^{2\nu},\pi^{3\nu})_{\nu=0,\dots,3}$
defined on $[0,T)\times \mathbb{R}^3$
such that 
$\Psi$ is a solution to (1)-(2) in  $[0,T)\times \mathbb{R}^3$ and 
$\Psi = \mathring{\Psi}$ on $\{ t = 0 \}$. Moreover, the solution $\Psi$ is causal in the sense
of Definition \ref{D:Causality}.
\end{theorem}

\noindent \emph{Proof of Theorem \ref{T:Existence}:} The calculations 
provided in Section I 
and in the proof of Theorem \ref{T:Sufficient} imply that, under the assumptions, 
the characteristic polynomial of the system evaluated at the initial data is a product of 
strictly hyperbolic polynomials. One also sees that intersection of the interior of the 
characteristic cones defined by these strictly hyperbolic polynomials has non-empty
interior and lies outside the light-cone defined by the metric. Under these
circumstances we can apply theorems A.18, A.19, and A.23 of 
\cite{DisconziFollowupBemficaNoronha} to conclude the result (the remaining
assumptions of these theorems are easily verified in our case). \hfill \qed

\bigskip

For the sake of brevity, we refer readers to \cite{RodinoGevreyBook} for a definition of $\mathcal{G}^\delta$,
making only the following remarks.
The case of $\delta =1$ corresponds to the space of analytic functions, of which 
$\mathcal{G}^\delta$ with $\delta > 1$ is a generalization. This is why $\mathcal{G}$ is sometimes
referred to as the space of quasi-analytic functions.
The usefulness of Gevrey functions to the study of hyperbolic problems is at least two-fold. On the one 
hand, one can prove very general existence and uniqueness theorems for Gevrey data given 
on a non-characteristic surface that are akin to the Cauchy-Kovalewskaya theorem for analytic data.
On the other hand, an advantage of Gevrey maps over analytic ones is that one can construct
Gevrey functions that are compactly supported; hence one can appeal to the type
of localization arguments that are so useful in the study of hyperbolic equations. This is 
particularly important when one is considering coupling to Einstein's equations.

While typical evolution problems consider solutions in more general function spaces than $\mathcal{G}^\delta$,
we stress that ours is the very first existence and uniqueness result for equations (1)-(2). In other words,
while it is desirable to extend our result to more general function spaces,  Theorem \ref{T:Existence} 
is important because it shows, for the very first time in the literature, that the initial value
problem for equation (1)-(2) is well-defined, so that it is meaningful to talk about 
solutions.

We remark that the diagonalization of $\pi_{\mu\nu}$ was carried out in terms of orthonormal frames which can be defined for any Lorentzian metric. Also, our computations are manifestly covariant. 
Thus, the result of Theorem \ref{T:Existence}
remains true in a general globally hyperbolic space-time, as mentioned in the main text.
This includes, in particular, the cases where the equations hold in a globally hyperbolic
subset of Minkowski space or in
 $I\times \mathbb{T}^3$ with the Minkowski metric,
where $I \subseteq \mathbb{R}$ is an interval and $\mathbb{T}^3$ is the three-dimensional torus.
Moreover, as also mentioned in the main text, the result extends to the case when (1)-(2) are coupled
to Einstein's equations. 
This follows by computing the characteristic determinant of the coupled
system and observing that it factors into the product of the characteristic determinant of (1)-(2), which we analyzed here,
and the characteristic determinant of Einstein's equations. The argument is the same as given in
\cite{BemficaDisconziNoronha_IS_bulk}.

\section{V. Insufficiency of conditions for causality\label{S:Insufficient}}
In this Section, we show that conditions (4), albeit necessary, are not sufficient for causality. 
We do this by showing that causality can be violated if we only assume (A1)-(A3) and (4).

Thus, suppose that (A1)-(A3) and (4) hold. Consider the case where 
(Ins1) $\delta_{\pi\pi}=\tau_{\pi\pi}/4$, $\delta_{\Pi\Pi}=0$, $\zeta+\lambda_{\Pi\pi}\Lambda_a\ge0$, $\frac{\lambda_{\Pi\pi}}{\tau_\Pi}+c_s^2-\frac{\tau_{\pi\pi}}{12\tau_\pi}>0$, and $1-c_s^2-\frac{\tau_{\pi\pi}}{3\tau_\pi}-\frac{\lambda_{\Pi\pi}}{\tau_\Pi}<0$. Also, the parameters as well as $c_s^2$ obey the necessary conditions (4). 
Assume also that (Ins2) $\Lambda_3=\Lambda_2>0$, i.e., $\Lambda_3$ is a degenerated eigenvalue. Then, we may write 
\be
G(k)=1-\sum_a\frac{R_a\hat{v}_a^2}{\bar{m}_a}-\sum_{\underset{a<b}{a,b}}\frac{S_{ab}\hat{v}_a^2\hat{v}_b^2}{\bar{m}_a\bar{m_b}},
\ee
where 
\be
R_a=\frac{1}{6\tau_\pi}\left [2\eta+\lambda_{\pi\Pi}\Pi+\frac{\tau_{\pi\pi}}{2}\Lambda_a\right ]+\frac{\zeta+\lambda_{\Pi\pi}\Lambda_a}{\tau_\Pi}+(\varepsilon+P+\Pi+\Lambda_a)c_s^2
\ee
and
\be
S_{ab}=\frac{\tau_{\pi\pi}}{6\tau_\pi}\left (\frac{\Lambda_{\Pi\pi}}{\tau_\Pi}+c_s^2-\frac{\tau_{\pi\pi}}{12\tau_\pi}\right )(\Lambda_a-\Lambda_b)^2.
\ee
From (4a) together with the above choices we have that $R_a,S_{ab}>0$. Now, let us define 
\be
\tilde{m}_a\equiv\varepsilon+P+\Pi+\Lambda_a-\frac{1}{2\tau_\pi}(2\eta+\lambda_{\pi\Pi}\Pi)-\frac{\tau_{\pi\pi}}{2\tau_\pi}\Lambda_a.
\ee 
Then, (4f) can be written as 
\be
\label{proof1-1}
\tilde{m}_d - R_d\ge 0,
\ee 
culminating into $\tilde{m}_d>0$. Note that this must hold for any $d=1,2,3$. Let us consider
the case where $a_1$ is such that $\tilde{m}_{a_1} - R_{a_1}=\min_{d}(\tilde{m}_d - R_d)$. Thus, if \eqref{proof1-1} is verified for $d=a_1$, it must be verified for all $d=1,2,3$. Now, we may choose the constraint in the parameters (Ins3) $\tilde{m}_{a_1} - R_{a_1}=0$, what is in accord with \eqref{proof1-1}. The remaining of this proof relies on the choice $\hat{v}_{a_1}=\sqrt{1-\epsilon^2}$, $\hat{v}_{a_2}=\epsilon$, and $\hat{v}_{a_3}=0$ for $\epsilon\in(0,1)$. The remaining of the proof relies on the assumption (Ins4) that if $a_1=3,2$, then $a_2=2,3$ while if $a_1=1$, then $a_2$ can be either 2 or 3. Thus, one can clearly see that
\bea
f(k)&=&\bar{m}_{a_3}\left (\bar{m}_{a_1}\bar{m}_{a_2}-\bar{m}_{a_1}R_{a_2}\epsilon^2-\bar{m}_{a_2} R_{a_1}(1-\epsilon^2)-S_{a_1a_2}\epsilon^2(1-\epsilon^2)\right ),\label{proof1}\\
\bar{m}_d&=&(\varepsilon+P+\Pi+\Lambda_d)k-\frac{1}{2\tau_\pi}(2\eta+\lambda_{\pi\Pi}\Pi)-\frac{\tau_{\pi\pi}}{4\tau_\pi}\left [\Lambda_d+\Lambda_{a_1}(1-\epsilon^2)+\Lambda_{a_2}\epsilon^2\right ]\nonumber\\
&=&\bar{m}_d^{0}-\frac{\tau_{\pi\pi}}{4\tau_\pi}(\Lambda_{a_2}-\Lambda_{a_1})\epsilon^2,
\eea
where we defined
\[\bar{m}_d^0\equiv(\varepsilon+P+\Pi+\Lambda_d)k-\frac{1}{2\tau_\pi}(2\eta+\lambda_{\pi\Pi}\Pi)-\frac{\tau_{\pi\pi}}{4\tau_\pi}\left (\Lambda_d+\Lambda_{a_1}\right ).\] 
From (4d) one may easily verify that $\bar{m}_d^0(k\ge1)\ge 0$. In particular,  
\be
\label{proof1-2}
\bar{m}_{a_1}^0(k=1)=\tilde{m}_{a_1}>0
\ee 
from \eqref{proof1-1}, while $\bar{m}_{a_2, a_3}^0(k=1)>\tilde{m}_{a_2, a_3}>0$. (Ins2) enables us to write (note that $a_2\ne 1$ according to (Ins4))
\be
\Lambda_{a_1}(1-\epsilon^2)+\Lambda_{a_2}\epsilon^2\begin{cases}
=\Lambda_3=\Lambda_2,\; \text{if}\; a_1=2, a_2=3\;\text{or}\;a_1=3, a_2=2,\\
<\Lambda_3, \;\text{if} \; a_1=1\;\forall\;\epsilon\in(0,1)
\end{cases},
\ee
what results into 
\be
\bar{m}_d\ge(\varepsilon+P+\Pi+\Lambda_d)k-\frac{1}{2\tau_\pi}(2\eta+\lambda_{\pi\Pi}\Pi)-\frac{\tau_{\pi\pi}}{4\tau_\pi}\left (\Lambda_d+\Lambda_3\right ),
\ee
and gives $\bar{m}_{d}(k\ge 1)\ge 0$ due to (4d) and $\bar{m}_{2,3}(k\ge 1)>0$ because $\tilde{m}_{d}>0$ from \eqref{proof1-1}.  

The roots of $f$ are the roots of $\bar{m}_{a_3}$ and the roots in the term in brackets in \eqref{proof1}. Let us define it as
\bea
\label{proof2-3}
\tilde{f}(k)&\equiv&\bar{m}_{a_1}\bar{m}_{a_2}-\bar{m}_{a_1}R_{a_2}\epsilon^2-\bar{m}_{a_2} R_{a_1}(1-\epsilon^2)-S_{a_1a_2}\epsilon^2(1-\epsilon^2)\nonumber\\
&=&\bar{m}_{a_1}\bar{m}_{a_2} G(k),
\eea 
where
\bea
\label{proof2}
G(k)=1-\frac{R_{a_2}\epsilon^2}{\bar{m}_{a_2}}-\frac{ R_{a_1}(1-\epsilon^2)}{\bar{m}_{a_1}}-\frac{S_{a_1a_2}\epsilon^2(1-\epsilon^2)}{\bar{m}_{a_1}\bar{m}_{a_2}}.
\eea
Note that since $\epsilon\in(0,1)$, the terms $\bar{m}_{a_1,a_2}(\bar{k})$ cannot be zero if $\bar{k}$ is a root of $\tilde{f}$ due to the term $S_{a_1a_2}$. Also, because $\tilde{f}(k)=(\rho+\Lambda_{a_1})(\rho+\Lambda_{a_2})k^2+\mathcal{O}(k)$ is a positive function after the greater real root due to (Ins3), then $\tilde{f}(k>1)>0$, or equivalently $G(k>1)>0$, guarantees that there is no real root for $k>1$. Because (Ins1) leads to $R_a,S_{ab}>0$ and since $\bar{m}_a(k>1)>\bar{m}_a(k=1)$, then condition $G(k>1)>0$ is equivalent to $G(k=1)\ge0$. In other words we must have that   
\bea
&&1-\frac{R_{a_2}\epsilon^2}{\bar{m}_{a_2}(k=1)}-\frac{ R_{a_1}(1-\epsilon^2)}{\bar{m}_{a_1}(k=1)}-\frac{S_{a_1a_2}\epsilon^2(1-\epsilon^2)}{\bar{m}_{a_1}(k=1)\bar{m}_{a_2}(k=1)}\ge0.\label{proof2-2}
\eea
Since $\epsilon<1$ we can expand \eqref{proof1-2} in powers of it and, after using \eqref{proof1-2} and (Ins3), obtain the causality condition
\bea
\label{proof3}
&&\left \{ 1-\frac{\tau_{\pi\pi}}{4\tau_\pi\tilde{m}_{a_1}}(\Lambda_{a_2}-\Lambda_{a_1})-\frac{R_{a_2}}{\bar{m}_{a_2}^0(k=1)}-\frac{S_{a_1a_2}}{\tilde{m}_{a_1}\bar{m}_{a_2}^0(k=1)}\right \}\epsilon^2+\mathcal{O}(\epsilon^4)\ge 0.
\eea
Now, by writing 
\[\bar{m}^0_{a_2}(k=1)=\tilde{m}_{a_1}+(\Lambda_{a_2}-\Lambda_{a_1})\left ( 1-\frac{\tau_{\pi\pi}}{4\tau_\pi}\right )\]
and
\[R_{a_2}=R_{a_1}+(\Lambda_{a_2}-\Lambda_{a_1})\left (c_s^2+\frac{\tau_{\pi\pi}}{12\tau_\pi}+\frac{\lambda_{\Pi\pi}}{\tau_\Pi}\right ),\]
and by means of (Ins3) we may rewrite 
\bea
\label{proof4}
&&1-\frac{R_{a_2}}{\bar{m}_{a_2}^0(k=1)}=\frac{\Lambda_{a_2}-\Lambda_{a_1}}{\bar{m}^0_{a_2}(k=1)}\left (1-c_s^2-\frac{\tau_{\pi\pi}}{3\tau_\pi}-\frac{\lambda_{\Pi\pi}}{\tau_\Pi}\right )\le 0.
\eea
Note that \eqref{proof4} is negative or zero because of (Ins2), (Ins3), and (Ins4). From (Ins2) and (Ins4), if $a_1=2,3$, then $a_2=3,2$ and $\Lambda_{a_2}-\Lambda_{a_1}=0$ while if $a_1=1$, then $a_2=2,3$ and $\Lambda_{a_2}-\Lambda_1>0$, resulting in $\Lambda_{a_2}-\Lambda_{a_1}\ge0$, while (Ins1) makes \eqref{proof4} negative or zero. As a consequence of \eqref{proof4}, the term proportional to $\epsilon^2$ in the LHS of \eqref{proof3} is negative and, for some small value of $\epsilon\in(0,1)$ it must become the leading term, turning the LHS of \eqref{proof3} strictly negative. Then, one concludes that the system is not causal and the necessary conditions (4) are not sufficient.

\section{VI. Formal definition of causality and conditions (C1) and (C2)\label{S:Formal}}

Since the notion of causality is central in our work, we find it appropriate to give its
precise mathematical definition. We also comment on how it is equivalent, 
in our context, to conditions (C1) and (C2).

Causality can be defined as follows (see  \cite[page 620]{ChoquetBruhatGRBook}  or \cite[Theorem 10.1.3]{WaldBookGR1984}  for more details).

\begin{definition}
\label{D:Causality}
Let $(\mathcal{M},g)$ be the Minkowski space.
Consider in $\mathcal{M}$ a system of partial differential equations for an unknown $\psi$, 
which we write as $\mathcal{P} \psi = 0$, where $\mathcal{P}$ is a differential 
operator (which is allowed to depend on $\psi$)\footnote{Since this is a system of PDEs, in coordinates it would be represented by 
equations of the form  $P^I_K \psi^K  = 0$, $I,K=1,\dots, N$, where 
$\{ \psi^K \}_{K=1}^N$ are local representations of $\psi$, e.g., the components
of $\psi$ if $\psi$ is a tensor, and 
$\mathcal{P}^I_K$ are differential operators (possibly depending on $\psi^K$).}.
Let $\varphi$ be a solution to the system.
We say that $\varphi$ is causal if
the following holds true: given a  Cauchy 
surface $\Sigma \subset \mathcal{M}$, 
for any point $x$ in the future of $\Sigma$, $\varphi(x)$ 
depends only on $\left.\varphi \right|_{J^-(x) \cap \Sigma}$, where $J^-(x)$ is the causal past of $x$.
\end{definition}

\begin{figure}[th]
\includegraphics[width=0.6\textwidth]{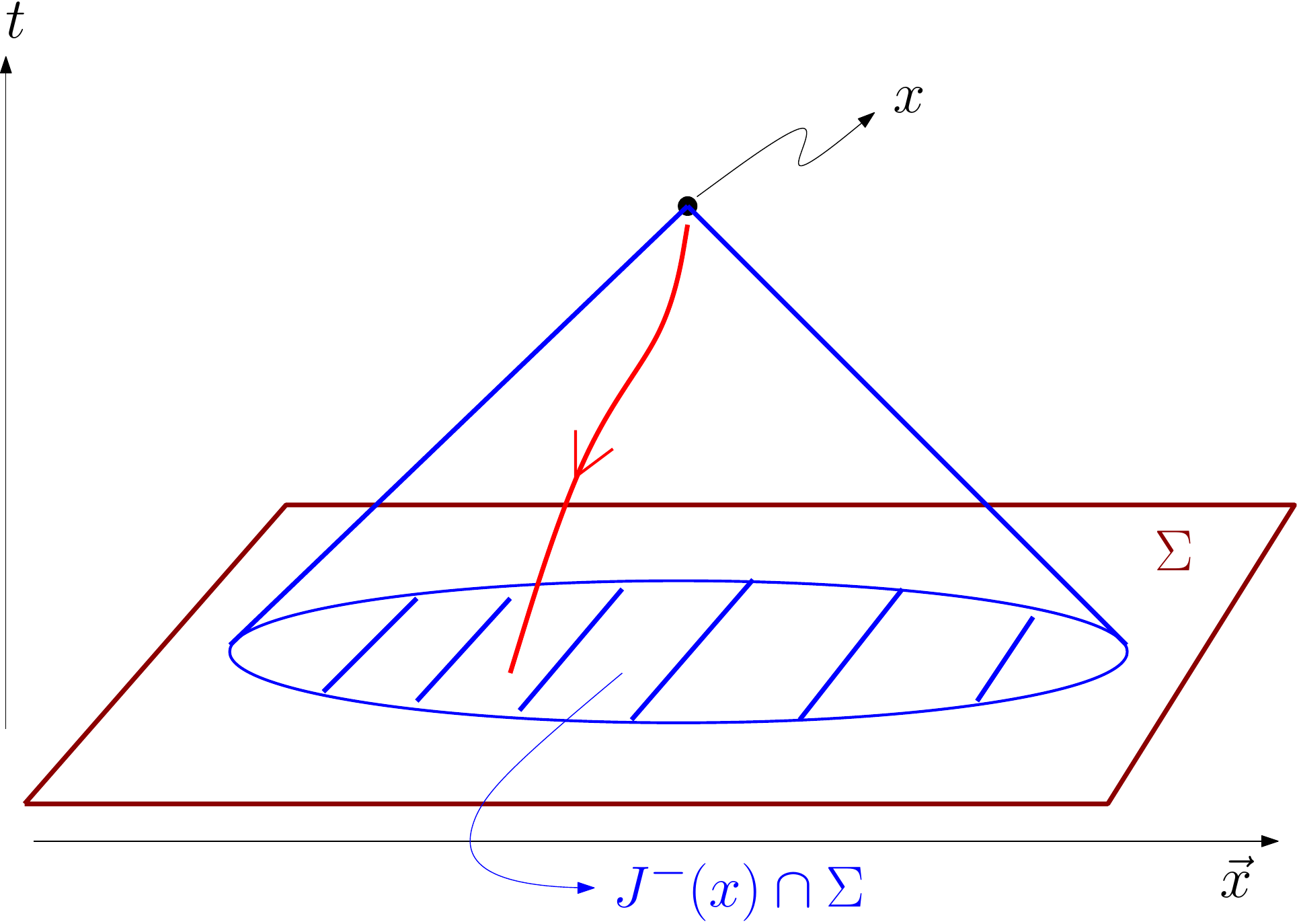}
\caption{(color online) Illustration of causality.  $J^-(x)$ 
is the past light-cone with vertex at $x$.
Points inside $J^-(x)$ can be joined to a point $x$ in space-time by a causal past directed curve (e.g. the red line). 
The value of $\varphi(x)$ depends only on $\left. \varphi \right|_{J^-(x) \cap \Sigma}$.
The Cauchy surface $\Sigma$ typically supports the initial data, in which case 
$\varphi(x)$ depends only on the initial data on $J^-(x) \cap \Sigma$.
}
\label{fig1}
\end{figure}

The case of most interest is when the Cauchy surface is the hypersurface 
$\{ t= 0 \}$ where initial data is prescribed. We also notice that since we are working in Minkowski
space, $J^-(x)$ is simply the past light-cone with vertex at $x$. The situation in 
Definition \ref{D:Causality} is illustrated in Fig.\ \ref{fig1}. 
In particular, causality implies that $\varphi(x)$ remains unchanged if the the values
of $\varphi$ along $\Sigma$ are altered\footnote{Causality can be equivalently stated in the following manner. 
If $\varphi_0$ and 
$\widetilde{\varphi}_0$ are two sets of initial data for the system 
prescribed on $\Sigma$ and 
such that 
$\varphi_0 = \widetilde{\varphi}_0$ on a subset $S \subset \Si$, and $\varphi$ and 
$\widetilde{\varphi}$ are the 
corresponding solutions to the equations, then $\varphi = \widetilde{\varphi}$ on $D_g^+(S)$, where 
$D^+_g(S)$ is the future domain of dependence of $S$ \cite[Theorem 10.1.3]{WaldBookGR1984}.}
 only outside $J^-(x) \cap \Sigma$. 
 Observe that this definition says that $\varphi(x)$ can only
be influenced by points in the past of $x$ that are causally connected to $x$, so no information
is allowed to propagate faster than the speed of light.

Definition \ref{D:Causality} is for a given solution 
$\varphi$ to the system. While it would be desirable to state causality as a general property of the 
system $\mathcal{P} \psi = 0$, i.e., saying that the system is causal if any solution
is causal in the sense of Definition \ref{D:Causality}, this would be too restrictive, as
it can be seen from our discussion of the equation $- \psi_{tt} + (1+\psi) \Delta \psi = 0$
in the Conclusion.

The connection between Definition \ref{D:Causality}
and conditions (C1) and (C2) is via the characteristics of the system $\mathcal{P} \psi = 0$.
It is beyond the scope of this Supplemental Material to provide a detailed description
of the connections between Definition \ref{D:Causality} and the system's characteristics.
We refer readers to Appendix A of \cite{DisconziFollowupBemficaNoronha},
\cite[Chapter VI]{Courant_and_Hilbert_book_2}, and \cite{Leray_book_hyperbolic}.
Here, we restrict ourselves to the following comments. Finite speed of propagation is a property
of hyperbolic equations. For such equations, there exist domains of dependence that show
precisely how the values of a solution at a point $x$ is determined solely by values within
a domain of dependence in the past with ``vertex" at $x$ (this is exactly the generalization of the
past light-cone).
The domain of dependence, in turn, is determined by the system's characteristics.
While it is mathematically possible for hyperbolic equations to exhibit domains of dependence
where information propagates faster than the speed of light (see, again, discussion in the 
Conclusion), for solutions to be causal (i.e., to not have faster-than-light signals), the domains
of dependence must always lie inside the light-cones. This is equivalent to 
the statement (C1) and (C2) that we have used.

Definition \ref{D:Causality} can be generalized to arbitrary globally hyperbolic spaces,
which is needed for the aforementioned generalization of our Theorems to this setting.
Again,
we refer to  Appendix A of \cite{DisconziFollowupBemficaNoronha},
\cite[Chapter VI]{Courant_and_Hilbert_book_2}, and \cite{Leray_book_hyperbolic}.

\bibliography{References.bib}

\end{document}